\documentclass[10pt,journal,twocolumn]{IEEEtran}

\IEEEoverridecommandlockouts
\usepackage{cite}
\usepackage{xcolor}
\usepackage{textcomp}
\usepackage{url}
\usepackage{verbatim}
\usepackage{enumerate}
\usepackage{amsmath,amssymb,amsfonts}
\usepackage{amsthm}
\usepackage{bm}
\usepackage{graphicx}
\usepackage{subcaption}
\usepackage{epstopdf}
\usepackage{array}
\usepackage{booktabs}
\usepackage{makecell}
\usepackage{adjustbox}
\usepackage{algorithm}
\usepackage{algorithmic}

\newtheoremstyle{mystyle}
{3pt}                 
{3pt}                 
{\normalfont}         
{}                    
{\bfseries\itshape}   
{.}                   
{0.5em}               
{\underline{#1}~{\upshape #2}\thmnote{ (#3)}}
\theoremstyle{mystyle}
\newtheorem{remark}{Remark}

\newtheorem{lemma}{Lemma}

\allowdisplaybreaks[4]

\def\BibTeX{{\rm B\kern-.05em{\sc i\kern-.025em b}%
\kern-.08em T\kern-.1667em\lower.7ex\hbox{E}%
\kern-.125emX}}

\begin{document}

\title{Fluid Antenna-Enabled Secure UAV Communication Systems Under Uncertain Eavesdropper Locations}

\author{
Wenchao Liu, Weijie Yuan, Jun Wu, Xuhui Zhang, Jinke Ren, and Changsheng You

\thanks{
Wenchao Liu, Weijie Yuan, and Jun Wu are with the School of Automation and Intelligent Manufacturing, Southern University of Science and Technology, Shenzhen 518055, China (e-mail: wc.liu@foxmail.com; yuanwj@sustech.edu.cn; wuj2021@mail.sustech.edu.cn).
}

\thanks{
Xuhui Zhang and Jinke Ren are with the School of Science and Engineering, Shenzhen Future Network of Intelligence Institute, and Guangdong Provincial Key Laboratory of Future Networks of Intelligence, The Chinese University of Hong Kong (Shenzhen), Longgang, Shenzhen, Guangdong 518172, China (e-mail: xu.hui.zhang@foxmail.com; jinkeren@cuhk.edu.cn).
}

\thanks{Changsheng You is with the Department of Electronic and Electrical Engineering, Southern University of Science and Technology, Shenzhen 518055, China (e-mail: youcs@sustech.edu.cn).
}

}

\maketitle

\begin{abstract}
Conventional uncrewed aerial vehicle (UAV)-enabled secure communication systems generally rely on fixed-position antennas (FPAs), which limit antenna-level spatial reconfigurability and the available degrees of freedom (DoFs) for physical-layer security (PLS) enhancement. To address this limitation, we investigate a fluid antenna (FA)-enabled UAV secure communication system in the presence of an eavesdropper whose exact location is unknown but confined to a prescribed uncertainty region. We adopt a Rician fading model to characterize the line-of-sight (LoS) and non-line-of-sight (NLoS) components of the air-to-ground (A2G) channels. Accounting for the transmission time consumed by FA reconfiguration, we formulate an expected accumulated secrecy throughput maximization problem by jointly optimizing the UAV trajectory, transmit beamforming, and FA positions over a finite mission period. To handle the eavesdropper location uncertainty, we adopt a sampling-based robust design that considers the worst-case accumulated eavesdropping throughput over a finite set of candidate locations. To address the resulting non-convex optimization problem, we develop an alternating optimization (AO)-based algorithm to obtain a high-quality suboptimal solution. We further discuss the algorithm initialization and analyze its convergence behavior and computational complexity. Simulation results demonstrate that the proposed joint design achieves higher accumulated secrecy throughput than the considered benchmark schemes, highlighting the benefits of coordinating UAV mobility with FA position reconfiguration for secure aerial communications.
\end{abstract}


\section{Introduction}
\IEEEPARstart{U}{ncrewed} aerial vehicles (UAVs), also known as drones or autonomous aerial platforms, have emerged as key enablers of next-generation wireless communication infrastructures. Their high mobility and flexible deployment provide significant advantages in scenarios where conventional terrestrial networks encounter inherent limitations \cite{11503149,11443232}. As the core components of low-altitude wireless networks (LAWNs), UAVs facilitate the formation of dynamic and reconfigurable aerial connectivity layers \cite{7888557,10693833}.
On one hand, UAVs can deliver on-demand and resilient air-to-ground (A2G) communication services in remote or infrastructure-deficient regions, where the deployment of terrestrial base stations (BSs) is often economically prohibitive \cite{9779853}. On the other hand, UAVs can act as temporary aerial BSs in high-traffic hotspot regions to complement overloaded ground networks and alleviate congestion \cite{9456851}.
In such deployments, UAVs typically establish line-of-sight (LoS)-dominant links with ground users, thereby improving spectral efficiency, enhancing throughput, and reducing end-to-end latency \cite{10098686}. Owing to these appealing advantages, UAV-enabled wireless communication has attracted significant attention from both academia and industry in recent years.

However, while the LoS channel condition enhances service quality, it also exposes UAV communication links to severe eavesdropping attacks due to the wide coverage and high predictability of aerial channels \cite{8254370}. This vulnerability highlights the importance of physical-layer security (PLS) as a crucial safeguard to ensure confidentiality. The main objective of PLS is to enhance the secrecy rate, defined as the rate difference between the legitimate user and the eavesdropper \cite{9905663}. In such open and dynamic wireless environments, PLS techniques can provide an additional layer of protection to complement conventional cryptographic mechanisms \cite{11214460}. By intentionally degrading the eavesdropper’s channel while maintaining or improving the quality of service for legitimate users, PLS enables a fundamental layer of protection \cite{8626132,10938968}. Accordingly, designing robust beamforming and resource allocation strategies that can adapt to the mobility of UAVs and potential eavesdroppers is essential for achieving secure and reliable UAV communications.

Although UAV mobility provides additional flexibility in adapting wireless channels, conventional fixed-position antennas (FPAs) have fixed array geometries and cannot exploit antenna-level position reconfiguration. This limits the spatial degrees of freedom (DoFs) available for jointly shaping the legitimate and eavesdropping links \cite{9264694,you2025next}. The inherent limitation in beamforming flexibility with static antenna architectures motivates the exploration of more adaptive transceiver technologies.
Recently, fluid antennas (FAs), also referred to as movable antennas (MAs), have emerged as a promising solution that allows radiating elements to dynamically adjust their positions within a conductive fluid medium \cite{10508218,10753482,10286328}. This capability introduces a new dimension of adaptability in wireless system design, particularly beneficial for UAV-enabled secure communications.
As UAVs traverse dynamic environments, FAs can reconfigure antenna positions in real time to optimize beamforming patterns \cite{10654366, 11240557}. This spatial reconfigurability enables the UAV to enhance the desired channel gains while suppressing signal leakage toward potential eavesdropper locations. However, FA reconfiguration incurs movement delay and reduces the time available for information transmission. Therefore, maximizing the accumulated secrecy throughput requires jointly considering the secrecy performance and the associated reconfiguration overhead.

Motivated by these considerations, we investigate an FA-enabled secure UAV communication system. Specifically, we aim to maximize the expected accumulated secrecy throughput of all legitimate users over the UAV mission by jointly optimizing the UAV trajectory, transmit beamforming, and FA positions, while accounting for FA reconfiguration overhead. The main contributions of this paper are summarized as follows: 
\begin{itemize}
\item We investigate an FA-enabled multiuser secure UAV communication system in the presence of a single eavesdropper with uncertain location. The UAV is equipped with an FA array, and a two-phase transmission protocol is considered to account for antenna reconfiguration and information transmission. Based on this model, we formulate an expected accumulated secrecy throughput maximization problem over a finite UAV mission.
\item We characterize the A2G channels using a Rician fading model that incorporates both LoS and non-line-of-sight (NLoS) components. To account for channel fading, we adopt the expected accumulated secrecy throughput as the optimization metric. A statistical CSI-based approximation is further employed to obtain tractable expressions for the legitimate and eavesdropping throughputs.
\item To handle eavesdropper location uncertainty, we develop a sampling-based robust optimization framework by generating $J$ candidate locations within the uncertainty region. For each legitimate user, the candidate location yielding the maximum accumulated eavesdropping throughput is considered in the optimization, thereby providing robustness against the worst-case candidate among the sampled locations.
\item To address the non-convexity arising from the coupled optimization variables, we introduce auxiliary variables and construct tractable surrogate subproblems. Based on these transformations, an AO-based algorithm is developed to iteratively optimize the UAV trajectory, transmit beamforming, and FA positions. The initialization, convergence behavior, and computational complexity of the proposed algorithm are also analyzed.
\end{itemize}
Numerical results validate the benefits of jointly optimizing UAV trajectory, transmit beamforming, and FA positions in improving the accumulated secrecy throughput.

The remainder of this paper is organized as follows. Section II reviews the related work. Section III presents the FA-enabled secure UAV communication system incorporating the FA technology and formulates the corresponding optimization problem. An AO-based algorithm is developed in Section IV, and its initialization, convergence, and computational complexity are analyzed. Numerical results are provided in Section V, and Section VI concludes the paper with summary remarks.

\section{Related Work}
With the increasing demand for ubiquitous wireless coverage and edge artificial intelligence (AI), UAV-enabled LAWNs have attracted considerable research attention. UAVs can enhance network coverage and communication capacity while facilitating efficient data collection through optimized trajectories \cite{9321340}. In multi-UAV systems, federated learning (FL)-based methods were proposed to jointly optimize trajectory planning and resource allocation \cite{10980172}. Moreover, UAV-enabled mobile edge computing (MEC) systems exploited UAV mobility to improve computational task performance, addressing the growing demand for edge AI \cite{9703683,10606316}. Recent works also explored the integration of UAVs with emerging technologies, including integrated sensing and communication (ISAC) \cite{9916163}, reconfigurable intelligent surfaces (RIS) \cite{9758649}, wireless FL \cite{10220154,10972043}, and large AI models \cite{11201912}, enabling synergistic gains in sensing, communication, computation, and edge intelligence.

Despite these benefits, ensuring secure data transmission and protecting legitimate users against eavesdropping in UAV-enabled LAWNs remain critical challenges. For example, secure data offloading in UAV-enabled MEC systems was enhanced using a jammer to improve secure computation capacity \cite{9450021}, while closed-form expressions for connection outage probability and secrecy outage probability were derived to characterize the reliability and security of multi-antenna UAV systems \cite{9970352}. UAV-relay-assisted maritime MEC systems were investigated to optimize secure computing capacity \cite{10310294}. Furthermore, combining UAVs with non-orthogonal multiple access (NOMA) improved both security and reliability \cite{10561528}, and RIS-assisted UAV networks maximized legitimate communication rates while suppressing eavesdropping through coordinated sensing \cite{wang2025coordinated}. These studies demonstrated that advanced techniques could significantly enhance the security of UAV-enabled LAWNs.

Beyond UAV systems, FAs have emerged as a promising technology for enhancing PLS in wireless networks. A fluid antenna system (FAS)-enabled PLS framework was proposed to analytically characterize secrecy performance over correlated fading channels \cite{10694739}, while low-complexity artificial-noise-aided secure transmission schemes were investigated for a single-antenna BS serving FA-equipped legitimate users \cite{10092780}. In rate-splitting multiple access (RSMA) and ISAC scenarios, FAs enabled joint beamforming and antenna positioning to simultaneously serve legitimate users and jam eavesdroppers under imperfect channel state information (CSI) \cite{zhang2025fluid}. Additionally, in NOMA networks, FAs could dynamically steer nulling beams toward eavesdroppers while enhancing legitimate reception, achieving significant secrecy gains over conventional FPA systems \cite{mai2024secure}. These studies demonstrate the potential of FAs for enhancing PLS in next-generation wireless networks.

Among the existing studies, \cite{11214460}, \cite{10654366}, and \cite{11156108} are particularly relevant to this work. Specifically, \cite{11214460} investigated the secrecy performance of UAV macro-mobility and MA micro-mobility using separate optimization frameworks, highlighting their respective advantages under different system conditions. In \cite{10654366}, the UAV trajectory, transmit beamforming, and MA positions were jointly optimized to maximize the achievable data rate in a multiuser UAV communication system. However, PLS and eavesdropper location uncertainty were not considered. In contrast, \cite{11156108} investigated MA-enabled secure transmission under uncertain eavesdropper locations by jointly optimizing transmit beamforming and MA positions at a BS, without considering UAV trajectory optimization. Unlike these studies, this paper investigates an FA-enabled multiuser secure UAV communication system that jointly optimizes the UAV trajectory, transmit beamforming, and FA positions under uncertain eavesdropper locations, while accounting for Rician A2G fading and FA movement delay in maximizing the accumulated secrecy throughput.

\section{System Model and Problem Formulation}
In this section, we first introduce the network model, channel model, and performance metric of the proposed secure communication system in detail. Subsequently, we formulate a joint optimization problem aimed at maximizing the accumulated secrecy throughput of the legitimate users.
 
\begin{figure}[t]
\centering
\includegraphics[width=1\linewidth]{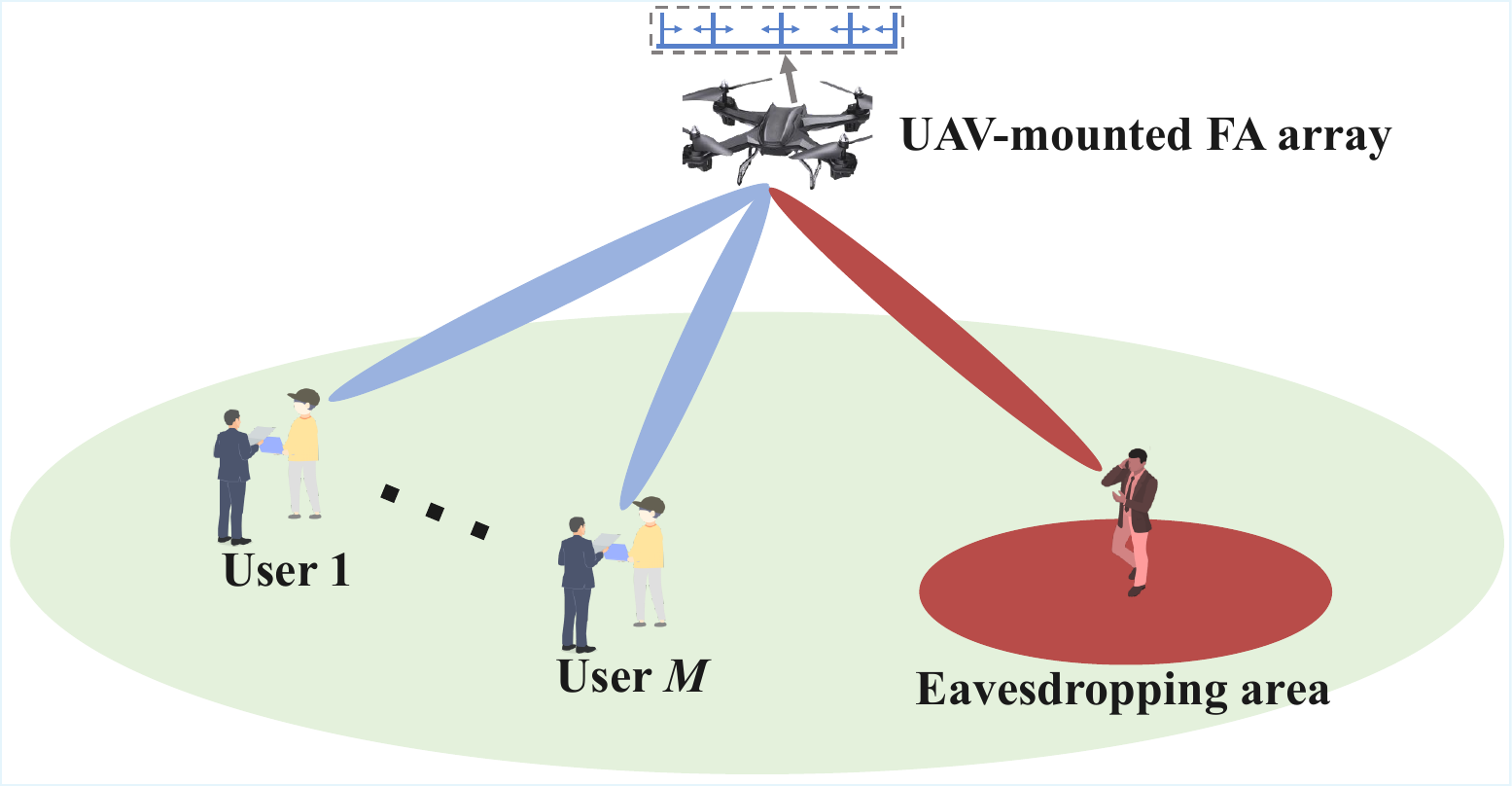}
\caption{Illustration of the considered FA-enabled secure UAV communication system with multiple users.}
\label{system model 1}
\end{figure}

\subsection{Network Model}
As illustrated in Fig.~\ref{system model 1}, we consider an FA-enabled secure UAV communication system. In this system, a UAV equipped with an FA array comprising $K$ antenna elements simultaneously transmits information signals to $M$ single-antenna ground legitimate users, while a single-antenna eavesdropper attempts to eavesdrop those signals.
Let $\mathcal{M} = \{1, \dots, M\}$ and $\mathcal{K} = \{1, \dots, K\}$ denote the sets of users and FAs. We assume that the UAV knows the locations of all legitimate users and the statistical channel information (CSI). In contrast, the eavesdropper’s exact location is unknown, and only an eavesdropping region $\mathcal{E}$ containing the eavesdropper is available \cite{11156108}. 

We consider a finite mission period $[0, T]$, which is uniformly discretized into $N$ time slots, each with a duration of $\tau = T/N$. Let $\mathcal{N} = \{1,\dots, N\}$ denote the set of time slots. The slot duration $\tau$ is chosen sufficiently small such that the UAV's location can be approximated as constant within each time slot. Throughout the mission, the UAV operates at a fixed altitude $H$, assumed to be sufficiently high to avoid potential blockages from ground obstacles or buildings.
Therefore, the three-dimensional (3D) locations of the user $m$, the eavesdropper, and the UAV in time slot $n$ are denoted as $(\mathbf{q}_{m}, 0)$, $(\mathbf{q}_{\rm{E}}, 0)$, and $(\mathbf{q}[n], H)$, respectively. Here, $\mathbf{q}_{m} = (x_{m}, y_{m})$ and $\mathbf{q}_{\rm{E}} = (x_{\rm{E}}, y_{\rm{E}}) \in \mathcal{E}$ denote the horizontal locations of user $m$ and the eavesdropper, while $\mathbf{q}[n] = (x[n], y[n])$ denotes the UAV’s horizontal location in time slot $n$. Moreover, $\mathbf{v}[n]=(\mathbf{q}[n]-\mathbf{q}[n-1])/\tau$ denotes the UAV's average horizontal velocity between two consecutive time slots.

\subsection{Channel Model}
In this system, the size of the FA array mounted on the UAV is significantly smaller than the signal propagation distance, ensuring far-field conditions between the UAV and each user \cite{11543384}. Under the far-field assumption, adjusting the FA positions mainly affects the phase responses of the propagation paths. In this work, the FA-position dependence is explicitly modeled through the deterministic LoS component, while the NLoS component is approximated using an i.i.d. complex Gaussian model.
Therefore, the channel vector from the UAV to the user $m$ and the channel vector from the UAV to the eavesdropper in time slot $n$ are denoted as $\mathbf{h}_{m}[n] \in \mathbb{C}^{K \times 1}$ and $\mathbf{h}_{\rm{E}}[n] \in \mathbb{C}^{K \times 1}$, characterized by the general Rician fading model, i.e., 
\begin{equation} 
\mathbf{h}_{\tilde{m}}[n] = \sqrt{ \frac{ \beta_{\tilde{m}}[n] \kappa_{\tilde{m}}[n] }{ \kappa_{\tilde{m}}[n] + 1 } } \mathbf{\overline{h}}_{\tilde{m}}[n] + \sqrt{ \frac{ \beta_{\tilde{m}}[n] }{ \kappa_{\tilde{m}}[n] + 1 } } \mathbf{\tilde{h}}_{\tilde{m}}[n],
\label{lab1}
\end{equation}
where $\tilde{m} \in \mathcal{M} \bigcup \{ \rm{E} \}$. 
The terms appearing in \eqref{lab1} are defined as follows:
\begin{itemize}
\item $\beta_{\tilde{m}}[n] = \tfrac{h_0}{d_{\tilde{m}}^{2}[n]}$ denotes the large-scale channel power gain, where $h_0$ is the channel power gain at a reference distance of $1$ meter and $d_{\tilde{m}}[n] = \sqrt{\|\mathbf{q}[n] - \mathbf{q}_{\tilde{m}} \|^{2}+H^2}$ is the distance between the UAV and the user/eavesdropper $\tilde{m}$ in time slot $n$.
\item $\kappa_{\tilde{m}}[n] = c_{1} \exp{( c_{2} \theta_{\tilde{m}}[n] )}$ denotes the Rician factor, characterizing the power ratio between the LoS and the NLoS components, where $c_{1}$ and $c_{2}$ are constant coefficients determined by the specific environment \cite{8698468} and $\theta_{\tilde{m}}[n] = \arcsin \left( \tfrac{H}{\sqrt{\|\mathbf{q}[n] - \mathbf{q}_{\tilde{m}} \|^{2}+H^2}} \right)$ denotes the vertical AoD from the UAV to the user/eavesdropper $\tilde{m}$ in time slot $n$.
\item $\mathbf{\overline{h}}_{\tilde{m}}[n] = \left[
{\mathrm{e}}^{{\mathrm{j}} \frac{2 \pi}{\lambda} x_{1}[n] \sin (\theta_{\tilde{m}}[n]) },  \ldots, 
{\mathrm{e}}^{{\mathrm{j}} \frac{2 \pi}{\lambda} x_{K}[n] \sin (\theta_{\tilde{m}}[n]) }
\right]^{\mathsf{T}} \in \mathbb{C}^{K \times 1}$ denotes the deterministic LoS component, where $x_{k}[n]$ is the position of FA $k$ in time slot $n$.
\item $\mathbf{\tilde{h}}_{\tilde{m}}[n] \in \mathbb{C}^{K \times 1}$ denotes the random NLoS component, whose elements are assumed to be independently and identically distributed (i.i.d.) complex Gaussian random variables with zero mean and unit variance, i.e., $\tilde{\mathbf{h}}_{\tilde{m}}[n] \sim \mathcal{CN}(0, \mathbf{I}_{K})$.
\end{itemize}

Let $s_{m}[n] \sim \mathcal{CN}(0,1)$ and $\mathbf{w}_{m}[n] \in \mathbb{C}^{K \times 1}$ denote the information signal for the user $m$ with normalized power and the corresponding transmit beamforming vector, respectively. The transmitted signal from the UAV in time slot $n$ is denoted as $\mathbf{x}_{t}[n] = \sum_{m=1}^{M} \mathbf{w}_{m}[n] s_{m}[n]$. Therefore, the received signal at the user/eavesdropper $\tilde{m}$ in time slot $n$ can be expressed as
\begin{align}
y_{\tilde{m}}[n] =  \sum_{i=1}^{M} \mathbf{h}_{\tilde{m}}^{\mathsf{H}}[n] \mathbf{w}_{i}[n] s_{i}[n] + \nu_{\tilde{m}}[n],
\end{align}
where $\nu_{\tilde{m}}[n] \sim \mathcal{CN}(0,\sigma^2_{\tilde{m}})$ is the additive white Gaussian noise (AWGN) at the user/eavesdropper $\tilde{m}$.
Then, the resulting signal-to-interference-plus-noise ratio (SINR) of the user $m$ in time slot $n$ is given by
\begin{equation} 
\gamma_{m}[n] = \frac{  \left\vert \mathbf{h}_{m}^{\mathsf{H}}[n]
 \mathbf{w}_{m}[n]  \right \vert^{2}  }
{  \sum\limits_{ i = 1, i \neq m }^{ \mathcal{M} } 
\left\vert  \mathbf{h}_{m}^{\mathsf{H}}[n] \mathbf{w}_{i}[n] \right \vert^{2} + \sigma^2_{m}},
\end{equation}
and the resulting SINR at the eavesdropper for eavesdropping the signal intended for user $m$ is given by
\begin{equation} 
\gamma_{m,{\rm{E}}}[n] = \frac{  \left\vert \mathbf{h}_{\rm{E}}^{\mathsf{H}}[n]
 \mathbf{w}_{m}[n]  \right \vert^{2}  }
{  \sum\limits_{ i = 1, i \neq m }^{ \mathcal{M} } 
\left\vert  \mathbf{h}_{\rm{E}}^{\mathsf{H}}[n] \mathbf{w}_{i}[n] \right \vert^{2} 
 + \sigma^2_{\rm{E}}}.
\end{equation}

\subsection{Performance Metrics}

\begin{figure}[t]
\centering
\includegraphics[width=0.9\linewidth]{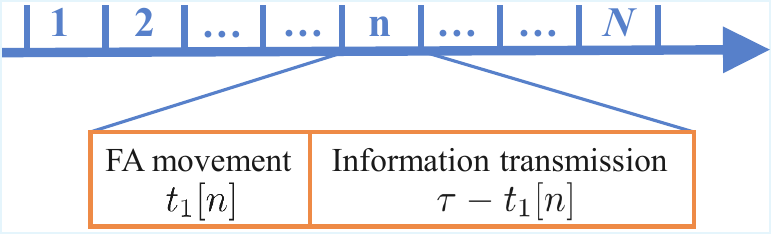}
\caption{Transmission protocol of the considered FA-enabled secure UAV communication system.}
\label{system model 2}
\end{figure}

During FA movement, rapid spatial variations of the channel and movement-induced Doppler effects may degrade communication quality. To avoid information transmission during FA reconfiguration, each time slot is divided into two phases, as illustrated in Fig. \ref{system model 2}. Specifically, the first phase, with duration $t_{1}[n]$, is allocated to FA movement, while the remaining duration $\tau-t_{1}[n]$ is used for information transmission \cite{11082461}. Taking the effective transmission duration into account, the amounts of information transmitted to user $m$, potentially intercepted by the eavesdropper for user $m$, and securely delivered to user $m$, all normalized by bandwidth, are respectively expressed as follows:
\begin{equation}
R_{m}[n] = (\tau - t_{1}[n]) \log_{2} ( 1 + \gamma_{m}[n] ),
\end{equation}
\begin{equation}
R_{m, {\rm{E}}}[n] = (\tau - t_{1}[n]) \log_{2} ( 1 + 
\gamma_{m,{\rm{E}}}[n] ),
\end{equation}
\begin{equation}
R_{m}^{\mathrm{sec}} [n] = \big[R_{m}[n] - R_{m, {\rm{E}}}[n] \big]^{+},
\end{equation}
where $[x]^{+} = \max\{x,0\}$. It is worth noting that  $R_m[n]$, $R_{m, {\rm{E}}}[n]$, and $R_{m}^{\mathrm{sec}} [n]$ are measured in bits/Hz, representing the information amounts per unit bandwidth accumulated during time slot $n$. Accordingly, summing the secrecy information over all users and time slots yields the accumulated secrecy throughput per unit bandwidth over the entire mission.

To characterize the secrecy performance over the entire mission under channel fading, we adopt the expected accumulated secrecy throughput of all legitimate users as the performance metric, which is expressed as
\begin{equation}
\mathrm{Objective:}\ \sum_{m=1}^{M}\sum_{n=1}^{N} \mathbb{E} 
\big[ R_{m}^{\mathrm{sec}}[n] \big].
\end{equation}

\subsection{Problem Formulation}
In this paper, we aim to maximize the expected accumulated secrecy throughput of the legitimate users over the entire mission by jointly optimizing the UAV trajectory $\{\mathbf{q}[n]\}$, the transmit beamforming $\{\mathbf{w}_{m}[n]\}$, and the FA positions $\{x_{k}[n]\}$. Mathematically, we can formulate the optimization problem as
\begin{subequations} 
\begin{flalign}
(\textbf{P1}):\  \max_{ \Phi_{1} } \,&  \sum_{m=1}^{M}\sum_{n=1}^{N} \mathbb{E} \big[ R_{m}^{\mathrm{sec}}[n] \big] \label{p1a}  \\
 {\rm{s.t.}}  \, &\mathbf{q}[1]= \mathbf{q}_{\rm{I}}, \quad 
 \mathbf{q}[N]=\mathbf{q}_{\rm{F}}, \label{p1b} \\
 & \| \mathbf{v}[n] \|  \le V_{\rm{max}}, \ \forall n\in \mathcal{N}\backslash\{1\}, \label{p1c} \\
 & \sum_{m= 1}^{M} \| \mathbf{w}_{m}[n] \|^2 \le P_{\max},\ \forall n, \label{p1d}  \\
& 0 \le x_{1}[n] \le \dots \le x_{K}[n] \le D_{\rm{FA}}, \ \forall n, \label{p1e} \\
& x_{k}[n] - x_{k-1}[n] \ge d_{\rm{min}}, \ \forall k\in \mathcal{K}\backslash\{1\}, \forall n, \label{p1f} \\
& \frac{ \left| x_{k}[n] - x_{k}[n-1] \right|  }{ \tilde{v} } \le t_{1}[n], \ \forall n \in \mathcal{N}\backslash\{1\}, \forall k, \label{p1g} 
\end{flalign} 
\end{subequations}
where $\Phi_{1} \triangleq \{ \mathbf{q}[n], \mathbf{w}_{m}[n], x_{k}[n] \}$, and $\tilde{v}$ in \eqref{p1g} denotes the FA movement speed. Constraint \eqref{p1b} ensures that the UAV starts from the initial location $\mathbf{q}_{\rm{I}}$ and ends at the final location $\mathbf{q}_{\rm{F}}$, thus satisfying the trajectory boundary conditions. Constraint \eqref{p1c} ensures that the UAV’s velocity does not exceed its maximum speed $V_{\max}$, ensuring kinematic feasibility during flight. Constraint \eqref{p1d} ensures that the total transmit power across all users in each time slot does not exceed the UAV’s maximum power budget $P_{\max}$.
Constraint \eqref{p1e} ensures that the FA positions are ordered and confined within the moving region $[0, D_{\rm FA}]$. Constraint \eqref{p1f} imposes a minimum separation distance $d_{\min}$ between adjacent FAs to prevent mutual coupling. Finally, constraint \eqref{p1g} ensures that the movement delay of the FA $k$ does not exceed $t_{1}[n]$ in time slot $n$.

\section{AO Algorithm For Solving Problem (\textbf{P1})}
Due to the strong coupling among the variables in the objective \eqref{p1a}, as well as the presence of the expectation operator $\mathbb{E}[\cdot]$, problem (\textbf{P1}) is challenging to solve directly. To address this challenge, we first employ a statistical CSI-based approximation to obtain tractable expressions for the ergodic rates. Based on these approximations, we formulate a deterministic surrogate optimization problem and then develop an AO-based algorithm to obtain a high-quality suboptimal solution.

\subsection{Problem Transformation}
\begin{lemma}
\rm{(\textit{Jensen’s Inequality}): For any convex function $f(\cdot)$ and random variable $x$ with finite expectation $\mathbb{E}[ x ]$, the following holds:}
\begin{equation}
\mathbb{E}\big[ f(x) \big]  \ge f \big(\mathbb{E}[ x ] \big).
\end{equation}
\label{lemma1}
\end{lemma}
In order to make the objective \eqref{p1a} easier to handle, by Lemma \ref{lemma1}, we can obtain
\begin{equation}
\mathbb{E} \big[ R_{m}^{\mathrm{sec}}[n] \big] \ge \bigg[ \mathbb{E} \big[ R_{m}[n] \big] - \mathbb{E} \big[ R_{m, {\rm{E}}}[n] \big] \bigg]^{+}.
\label{lab5}
\end{equation}

Note that the exact evaluation of the ergodic rates involves the expectation of logarithmic functions of random SINRs, which is generally difficult to handle in the subsequent optimization. To address this difficulty, following the statistical CSI-based rate approximation approaches in \cite{9500188,9034493,9355404}, we approximate the ergodic rates by replacing the instantaneous desired signal and interference powers with their corresponding average values. Accordingly, the approximate expressions for $\mathbb{E} [ R_{m}[n] ]$ and $\mathbb{E} \big[ R_{m, {\rm{E}}}[n] \big]$ are given by \eqref{lab2} and \eqref{lab3}, respectively. In these expressions, $\zeta^{\rm{los}}_{\tilde{m}}[n] = \frac{ \kappa_{\tilde{m}}[n] \beta_{\tilde{m}}[n] }{ \kappa_{\tilde{m}}[n] + 1 }$ and $\zeta^{\rm{nlos}}_{\tilde{m}}[n] = \frac{ \beta_{\tilde{m}}[n] }{ \kappa_{\tilde{m}}[n] + 1 }$, where $\tilde{m} \in \mathcal{M} \bigcup \{ \rm{E} \}$. The detailed derivations are provided in Appendix \ref{proof1}.

\begin{figure*}[!t]
\vspace*{-\baselineskip} 
\begin{equation}
\tilde{R}_{m}[n] = \left( \tau - t_{1}[n] \right) \log_{2} \left( 1 + \frac{ \left( \zeta^{\rm{los}}_{m}[n] \left| \mathbf{\overline{h}}_{m}^{\mathsf{H}}[n]\mathbf{w}_{m}[n] \right|^{2} + \zeta^{\rm{nlos}}_{m}[n] \| \mathbf{w}_{m}[n] \|^{2} \right) }
{  \sum\limits_{ i = 1, i \neq m }^{ \mathcal{M} } 
\left( \zeta^{\rm{los}}_{m}[n] \left| \mathbf{\overline{h}}_{m}^{\mathsf{H}}[n]\mathbf{w}_{i}[n] \right|^{2} + \zeta^{\rm{nlos}}_{m}[n] \| \mathbf{w}_{i}[n] \|^{2} \right) + \sigma^2_{m}} \right) ,
\label{lab2}
\end{equation}
\begin{equation}
\tilde{R}_{m, {\rm{E}}}[n] = \left( \tau - t_{1}[n] \right) \log_{2} \left( 1 + \frac{   \left( \zeta^{\rm{los}}_{\rm{E}}[n] \left| \mathbf{\overline{h}}_{\rm{E}}^{\mathsf{H}}[n]\mathbf{w}_{m}[n] \right|^{2} + \zeta^{\rm{nlos}}_{\rm{E}}[n] \| \mathbf{w}_{m}[n] \|^{2} \right) }
{  \sum\limits_{ i = 1, i \neq m }^{ \mathcal{M} } 
\left( \zeta^{\rm{los}}_{\rm{E}}[n] \left| \mathbf{\overline{h}}_{\rm{E}}^{\mathsf{H}}[n]\mathbf{w}_{i}[n] \right|^{2} + \zeta^{\rm{nlos}}_{\rm{E}}[n] \| \mathbf{w}_{i}[n] \|^{2} \right) + \sigma^2_{\rm{E}}} \right) .
\label{lab3}
\end{equation}
\hrulefill
\end{figure*}

Moreover, the UAV only knows the region in which the eavesdropper is located and cannot determine its exact location. Ideally, the worst-case secrecy performance should be evaluated over all possible eavesdropper locations within the continuous uncertainty region $\mathcal{E}$, which introduces infinitely many location-dependent constraints and significantly increases the computational difficulty. To address this challenge, we adopt a sampling-based robust optimization approach to approximate the continuous uncertainty region using a finite number of candidate locations \cite{8501983}. Specifically, $J$ candidate locations are independently and uniformly sampled within $\mathcal{E}$, and the sampled locations remain fixed throughout the optimization process. For each legitimate user, the candidate location yielding the maximum accumulated eavesdropping throughput over the mission is selected to characterize the corresponding worst-case scenario among the sampled locations. It is worth noting that the proposed design accounts for the worst-case eavesdropping performance among the sampled locations under the adopted statistical channel approximation. Let $\mathcal J=\{1,\ldots,J\}$ denote the set of sampled eavesdropper locations. Then, we have
\begin{equation}
 - \sum_{n=1}^{N} \tilde{R}_{m,j}[n] \ge \eta_{m}, \, \forall j, \forall m, \label{lab4}
\end{equation}
where $\eta_{m}$ denotes the introduced auxiliary variable. Based on the statistical rate approximations in \eqref{lab2} and \eqref{lab3}, together with the finite-location representation of the eavesdropping region, we obtain the following deterministic surrogate problem
\begin{subequations} 
\begin{flalign}
(\textbf{P2}):\ \max_{ \Phi_{2} } \, & \sum_{m=1}^{M}\sum_{n=1}^{N}  \tilde{R}_{m}[n] + \sum_{m=1}^{M} \eta_{m}  \label{p2a}  \\
 {\rm{s.t.}}  \, &\eqref{p1b}-\eqref{p1g}, \eqref{lab4}, \nonumber
\end{flalign} 
\end{subequations}
where $\Phi_{2} \triangleq \{ \eta_{m}, \mathbf{q}[n], \mathbf{w}_{m}[n], x_{k}[n]\}$. 

\begin{remark}
According to Jensen's inequality, the ergodic secrecy rate satisfies the following rigorous relationship, i.e.,
\begin{align}
\mathbb{E} \big[ R_{m}^{\mathrm{sec}}[n] \big] &\ge \bigg[ \mathbb{E} \big[ R_{m}[n] \big] - \mathbb{E} \big[ R_{m, {\rm{E}}}[n] \big] \bigg]^{+} \nonumber \\
& \ge \mathbb{E} \big[ R_{m}[n] \big] - \mathbb{E} \big[ R_{m, {\rm{E}}}[n] \big] \nonumber \\
& \approx \tilde{R}_{m}[n] - \tilde{R}_{m, {\rm{E}}}[n].
\label{lab6}
\end{align}
As discussed above, $\tilde{R}_{m}[n]$ and $\tilde{R}_{m, {\rm{E}}}[n]$ in \eqref{lab2} and \eqref{lab3} are statistical approximations of the corresponding ergodic rates. Therefore, their difference provides a tractable approximation of the rate difference in \eqref{lab6}. Furthermore, according to \eqref{lab4}, we have
\begin{align}
\sum_{m=1}^{M}\sum_{n=1}^{N} \mathbb{E} \big[ R_{m}^{\mathrm{sec}}[n] \big] &\ge \sum_{m=1}^{M}\sum_{n=1}^{N} \big( \tilde{R}_{m}[n] - \tilde{R}_{m, {\rm{E}}}[n] \big) \nonumber \\
& \approx \sum_{m=1}^{M}\sum_{n=1}^{N}  \tilde{R}_{m}[n] - \sum_{m=1}^{M} \max_{j\in\mathcal{J}} \sum_{n=1}^{N} \tilde{R}_{m,j}[n] \nonumber \\
& \ge \sum_{m=1}^{M}\sum_{n=1}^{N}  \tilde{R}_{m}[n] + \sum_{m=1}^{M} \eta_{m}.
\end{align}
Hence, the objective function of problem (\textbf{P2}) provides a tractable lower-bound representation of the finite-sample surrogate objective. Moreover, since the original constraints \eqref{p1b}-\eqref{p1g} remain unchanged, any feasible solution to problem (\textbf{P2}) satisfies the physical feasibility constraints of problem (\textbf{P1}).
\end{remark} 

Moreover, problem (\textbf{P2}) remains non-convex. To facilitate tractable optimization, we introduce an auxiliary variable $\omega_{m}[n]$ satisfying
\begin{equation}
1 + \frac{A_{m,m}[n]}{ \sum\limits_{ i = 1, i \neq m }^{ \mathcal{M} } A_{m,i}[n] + \sigma^2_{m} } \geq  \omega_{m}[n],
\label{p2b}
\end{equation}
where $A_{m,i}[n] = \zeta^{\rm{los}}_{m}[n] \left| \mathbf{\overline{h}}_{m}^{\mathsf{H}}[n]\mathbf{w}_{i}[n] \right|^{2} + \zeta^{\rm{nlos}}_{m}[n] \| \mathbf{w}_{i}[n] \|^{2}$ for notational simplicity. Therefore, the lower bound for the term $\tilde{R}_{m}[n]$ in \eqref{p2a} can be given by
\begin{equation}
\tilde{R}_{m}^{\rm{lb}}[n] = (\tau - t_{1}[n]) \log_{2}(\omega_{m}[n]).
\end{equation}

It can be seen from \eqref{p2b} that the left-hand side (LHS) exhibits a fractional structure, with both the numerator and denominator involving variables, rendering constraint \eqref{p2b} non-convex. Again, we introduce an auxiliary variable $\varpi_{m}[n]$, where
\begin{equation}
\varpi_{m}[n] \geq \sum\limits_{ i = 1, i \neq m }^{ \mathcal{M} } A_{m,i}[n] + \sigma^2_{m}.
\label{p2c}
\end{equation}
Therefore, constraint \eqref{p2b} can be rewritten as
\begin{equation}
\sum\limits_{ i = 1 }^{ \mathcal{M} } A_{m,i}[n] + \sigma^2_{m} \geq \omega_{m}[n] \varpi_{m}[n].
\label{p2d}
\end{equation}

\begin{lemma} 
\label{lemma2}
Given any two positive terms $x_{i}$ and $x_{j}$, the upper bound and lower bound for $x_{i}x_{j}$ can be expressed as
\begin{equation}
x_{i} x_{j} \leq \frac{1}{2} \left( \frac{x_{j}^{(l)}}{x_{i}^{(l)}} x_{i}^2 + \frac{x_{i}^{(l)}}{x_{j}^{(l)}} x_{j}^2 \right),
\end{equation}
\begin{equation}
x_{i} x_{j} \geq (x_{i}^{(l)} x_{j}^{(l)}) \left( 1 + \ln x_{i} + \ln x_{j} - \ln x_{i}^{(l)} - \ln x_{j}^{(l)} \right),
\end{equation}
where $x_{i}^{(l)}$ and $x_{j}^{(l)}$ are the given local points in the $l$-th iteration.
\begin{proof}
	Please refer to \cite{7547360}.
\end{proof}
\end{lemma}

According to Lemma \ref{lemma2}, we can obtain
\begin{align}
\omega_{m}[n] \varpi_{m}[n] \leq \frac{1}{2} \left( \frac{\omega_{m}^{(l)}[n]}{\varpi_{m}^{(l)}[n]} \varpi_{m}^{2}[n] + \frac{\varpi_{m}^{(l)}[n]}{\omega_{m}^{(l)}[n]} \omega_{m}^{2}[n] \right),
\end{align}
where $\omega_{m}^{(l)}[n]$ and $\varpi_{m}^{(l)}[n]$ are the local points in the $l$-th iteration. Therefore, constraint \eqref{p2d} can be reformulated as
\begin{align}
\sum\limits_{ i = 1 }^{ \mathcal{M} } A_{m,i}[n] + \sigma^2_{m} \geq \frac{1}{2} \left( \frac{\omega_{m}^{(l)}[n]}{\varpi_{m}^{(l)}[n]} \varpi_{m}^{2}[n] + \frac{\varpi_{m}^{(l)}[n]}{\omega_{m}^{(l)}[n]} \omega_{m}^{2}[n] \right). 
\label{p2e}
\end{align}

Similarly, a lower bound for the negative approximated eavesdropping throughput, $\{ - \tilde{R}_{m,j}[n] \}$, can be constructed as
\begin{equation}
\tilde{R}_{m,j}^{\rm{lb}}[n] = (\tau - t_{1}[n]) \log_{2}(\tilde{\omega}_{m,j}[n]),
\end{equation}
along with the associated constraints
\begin{equation} \small
\sum\limits_{ i = 1,i \neq m }^{ \mathcal{M} } B_{j,i}[n] + \sigma^2_{j} \geq \frac{1}{2} \left( \frac{\tilde{\omega}_{m,j}^{(l)}[n]}{\tilde{\varpi}_{j}^{(l)}[n]} \tilde{\varpi}_{j}^{2}[n] + \frac{\tilde{\varpi}_{j}^{(l)}[n]}{\tilde{\omega}_{m,j}^{(l)}[n]} \tilde{\omega}_{m,j}^{2}[n] \right),
\label{p2f}
\end{equation}
\begin{equation}
\tilde{\varpi}_{j}[n] \geq \sum\limits_{ i = 1 }^{ \mathcal{M} } B_{j,i}[n] + \sigma^2_{j},
\label{p2g}
\end{equation}
where $B_{j,i}[n] = \zeta^{\rm{los}}_{j}[n] \left| \mathbf{\overline{h}}_{j}^{\mathsf{H}}[n]\mathbf{w}_{i}[n] \right|^{2} + \zeta^{\rm{nlos}}_{j}[n] \| \mathbf{w}_{i}[n] \|^{2}$.
The terms $\tilde{\omega}_{m,j}[n]$ and $\tilde{\varpi}_{j}[n]$ are the introduced auxiliary variables, while $\tilde{\omega}_{m,j}^{(l)}[n]$ and $\tilde{\varpi}_{j}^{(l)}[n]$ denote the given local points in the $l$-th iteration.
Then, constraint \eqref{lab4} can be reformulated as
\begin{equation}
\sum_{n=1}^{N} \tilde{R}_{m,j}^{\rm{lb}}[n] \ge \eta_{m}. \label{p2h}
\end{equation}

Finally, the problem (\textbf{P2}) can be reformulated as
\begin{subequations} 
\begin{flalign}
(\textbf{P3}):\ \max_{ \Phi_{3} } \, & \sum_{m=1}^{M}\sum_{n=1}^{N}  \tilde{R}_{m}^{\rm{lb}}[n] + \sum_{m=1}^{M} \eta_{m}  \label{p3a}  \\
{\rm{s.t.}}  \, &\eqref{p1b}-\eqref{p1g}, \eqref{p2c}, \eqref{p2e}, \eqref{p2f}-\eqref{p2h}, \nonumber
\end{flalign} 
\end{subequations}
where $\Phi_{3} \triangleq \{ \eta_{m}, \mathbf{q}[n], \mathbf{w}_{m}[n], x_{k}[n], \omega_{m}[n], \varpi_{m}[n], \tilde{\omega}_{m,j}[n],$
$\tilde{\varpi}_{j}[n]\}$. 
In the following subsection, we alternately optimize the UAV trajectory $\{\mathbf{q}[n]\}$, the transmit beamforming $\{\mathbf{w}_{m}[n]\}$, and the FA positions $\{x_{k}[n]\}$, by fixing a set of variables while optimizing others.

\subsection{UAV Trajectory Optimization}
Given $\{ \mathbf{w}_{m}[n], x_{k}[n] \}$, the subproblem of (\textbf{P3}) for optimizing the UAV trajectory $\{\mathbf{q}[n]\}$ can be expressed as 
\begin{subequations} 
\begin{flalign}
(\textbf{P4}):\ \max_{ \Phi_{4} } \, & \sum_{m=1}^{M}\sum_{n=1}^{N}  \tilde{R}_{m}^{\rm{lb}}[n] + \sum_{m=1}^{M} \eta_{m}  \label{p4a}  \\
 {\rm{s.t.}}  \, &\eqref{p1b}-\eqref{p1c}, \eqref{p2c}, \eqref{p2e}, \eqref{p2f}-\eqref{p2h}, \nonumber
\end{flalign} 
\end{subequations}
where $\Phi_{4} \triangleq \{ \eta_{m}, \mathbf{q}[n], \omega_{m}[n], \varpi_{m}[n], \tilde{\omega}_{m,j}[n], \tilde{\varpi}_{j}[n]\}$. 
It can be seen from \eqref{lab1} that both the steering vector $\overline{\mathbf{h}}_{m}[n]$ and the Rician factor $\kappa_{m}[n]$ are highly nonlinear functions of $\mathbf{q}[n]$. To enhance the tractability of the trajectory optimization, we employ an iterative approximation strategy. Specifically, in the $(l+1)$-th iteration, $\overline{\mathbf{h}}_{m}[n]$ and $\kappa_{m}[n]$ are approximated based on $\mathbf{q}^{(l)}[n]$ obtained in the $l$-th iteration. Consequently, their approximated expressions can be written as
\begin{equation}
\overline{\mathbf{h}}_{m}^{(l)}[n] = \left[
{\mathrm{e}}^{{\mathrm{j}} \frac{2 \pi}{\lambda} x_{1}[n] \sin (\theta_{m}^{(l)}[n]) },  \ldots, 
{\mathrm{e}}^{{\mathrm{j}} \frac{2 \pi}{\lambda} x_{K}[n] \sin (\theta_{m}^{(l)}[n]) }
\right]^{\mathsf{T}}, 
\end{equation}
\begin{equation}
\kappa_{m}^{(l)}[n] = c_{1} \exp{( c_{2} \theta_{m}^{(l)}[n]  )}, 
\end{equation}
where $\theta_{m}^{(l)}[n] = \arcsin \left( \frac{H}{\sqrt{\|\mathbf{q}^{(l)}[n] - \mathbf{q}_{m} \|^{2}+H^2}} \right)$.
Therefore, the term $A_{m,i}[n]$ can be rewritten as
\begin{align}
A_{m,i}^{(l)}[n] &= \zeta^{\rm{los}}_{m}[n] \left| \left( \mathbf{\overline{h}}_{m}^{(l)}[n] \right)^{\mathsf{H}} \mathbf{w}_{i}[n] \right|^{2} + \zeta^{\rm{nlos}}_{m}[n] \| \mathbf{w}_{i}[n] \|^{2} \nonumber \\
&= d_{m}^{-2}[n] \widehat{A}_{m,i}^{(l)}[n],
\end{align}
where
\begin{equation}
\widehat{A}_{m,i}^{(l)}[n] =  \frac{  \kappa_{m}^{(l)}[n] h_0 }{ \kappa_{m}^{(l)}[n] + 1 } \left| \left( \mathbf{\overline{h}}_{m}^{(l)}[n] \right)^{\mathsf{H}}  \mathbf{w}_{i}[n] \right|^{2}
+ \frac{  h_0 \| \mathbf{w}_{i}[n] \|^{2} }{ \kappa_{m}^{(l)}[n] + 1 }.
\end{equation}

Obviously, $A_{m,i}^{(l)}[n]$ is non-concave with respect to $\mathbf{q}[n]$, but it is convex with respect to $d_{m}^{2}[n]$. Since the first-order Taylor expansion provides a global lower bound for a convex function, we can obtain the following lower bound for $A_{m,i}^{(l)}[n]$ at the given point $\left( d_{m}^{(l)}[n] \right)^{2}$, i.e.,
\begin{equation}
A_{m,i}^{(l)}[n] \geq \widehat{A}_{m,i}^{(l)}[n] \left( \frac{1}{\left( d_{m}^{(l)}[n] \right)^{2}} - \frac{d_{m}^{2}[n] - \left( d_{m}^{(l)}[n] \right)^{2}}{\left( d_{m}^{(l)}[n] \right)^{4}} \right).
\end{equation}
Moreover, by introducing an auxiliary variable $\xi_{m}[n]$, we can obtain the upper bound for $A_{m,i}^{(l)}[n]$, i.e.,
\begin{equation}
A_{m,i}^{(l)}[n] \leq {\mathrm{e}}^{\xi_{m}[n]} \widehat{A}_{m,i}^{(l)}[n],
\end{equation}
where $\xi_{m}[n]$ satisfies
\begin{equation}
{\mathrm{e}}^{-\xi_{m}[n]} \le \left(d_{m}^{(l)}[n] \right)^{2} + 2\left(\mathbf{q}^{(l)}[n] - \mathbf{q}_{m}\right) \left(\mathbf{q}[n] - \mathbf{q}^{(l)}[n]\right)^{\mathsf{T}}. 
\label{p4b}
\end{equation}
Therefore, constraints \eqref{p2c} and \eqref{p2e} can be transformed into
\begin{equation}
\varpi_{m}[n] \geq \sum\limits_{ i = 1, i \neq m }^{ \mathcal{M} } {\mathrm{e}}^{\xi_{m}[n]} \widehat{A}_{m,i}^{(l)}[n] + \sigma^2_{m}, 
\label{p4c}
\end{equation}
\begin{align}
\sum\limits_{ i = 1 }^{ \mathcal{M} } \widehat{A}_{m,i}^{(l)}[n] & \left( \frac{1}{\left( d_{m}^{(l)}[n] \right)^{2}} - \frac{d_{m}^{2}[n] - \left( d_{m}^{(l)}[n] \right)^{2}}{\left( d_{m}^{(l)}[n] \right)^{4}} \right) + \sigma^2_{m} \nonumber \\
&\geq \frac{1}{2} \left( \frac{\omega_{m}^{(l)}[n]}{\varpi_{m}^{(l)}[n]} \varpi_{m}^{2}[n] + \frac{\varpi_{m}^{(l)}[n]}{\omega_{m}^{(l)}[n]} \omega_{m}^{2}[n] \right). 
\label{p4d}
\end{align}

Similarly, we can transform \eqref{p2f} and \eqref{p2g} into the following forms
\begin{align}
\sum\limits_{ i=1,i\neq m }^{ \mathcal{M}}  \widehat{B}_{j,i}^{(l)}[n] & \left( \frac{1}{\left( d_{j}^{(l)}[n] \right)^{2}} - \frac{d_{j}^{2}[n] - \left( d_{j}^{(l)}[n] \right)^{2}}{\left( d_{j}^{(l)}[n] \right)^{4}} \right) + \sigma^2_{j} \nonumber \\
&\geq \frac{1}{2} \left( \frac{\tilde{\omega}_{m,j}^{(l)}[n]}{\tilde{\varpi}_{j}^{(l)}[n]} \tilde{\varpi}_{j}^{2}[n] + \frac{\tilde{\varpi}_{j}^{(l)}[n]}{\tilde{\omega}_{m,j}^{(l)}[n]} \tilde{\omega}_{m,j}^{2}[n] \right),
\label{p4e}
\end{align}
\begin{equation}
\tilde{\varpi}_{j}[n] \geq \sum\limits_{ i = 1 }^{ \mathcal{M} } 
{\mathrm{e}}^{\tilde{\xi}_{j}[n]} \widehat{B}_{j,i}^{(l)}[n] + \sigma^2_{j},
\label{p4f}
\end{equation}
where $\tilde{\xi}_{j}[n]$ is an introduced auxiliary variable and satisfies
\begin{equation}
{\mathrm{e}}^{-\tilde{\xi}_{j}[n]} \le \left(d_{j}^{(l)}[n] \right)^{2} + 2\left(\mathbf{q}^{(l)}[n] - \mathbf{q}_{j}\right) \left(\mathbf{q}[n] - \mathbf{q}^{(l)}[n]\right)^{\mathsf{T}}. 
\label{p4g}
\end{equation}

With the above transformations, problem (\textbf{P4}) can be reformulated as the following problem
\begin{subequations} 
\begin{flalign}
(\textbf{P5}):\ \max_{ \Phi_{5} } \, & \sum_{m=1}^{M}\sum_{n=1}^{N}  \tilde{R}_{m}^{\rm{lb}}[n] + \sum_{m=1}^{M} \eta_{m}  \\
 {\rm{s.t.}}  \, &\eqref{p1b}-\eqref{p1c}, \eqref{p2h}, \eqref{p4b}-\eqref{p4g},  \nonumber
\end{flalign} 
\end{subequations}
where $\Phi_{5} \triangleq \{ \eta_{m}, \mathbf{q}[n], \omega_{m}[n], \varpi_{m}[n], \tilde{\omega}_{m,j}[n], \tilde{\varpi}_{j}[n], \xi_{m}[n],$
$\tilde{\xi}_{j}[n]\}$. 
Obviously, problem (\textbf{P5}) is a standard convex optimization problem, which can be efficiently solved via CVX \cite{grant2008cvx}.

\subsection{Transmit Beamforming Optimization}
Given $\{ \mathbf{q}[n], x_{k}[n] \}$, the subproblem of (\textbf{P3}) for optimizing the transmit beamforming $\{ \mathbf{w}_{m}[n] \}$ can be expressed as
\begin{subequations} 
\begin{flalign}
(\textbf{P6}):\ \max_{ \Phi_{6} } \, & \sum_{m=1}^{M}\sum_{n=1}^{N}  \tilde{R}_{m}^{\rm{lb}}[n] + \sum_{m=1}^{M} \eta_{m}   \\
{\rm{s.t.}}  \, &\eqref{p1d}, \eqref{p2c}, \eqref{p2e}, \eqref{p2f}-\eqref{p2h}, \nonumber
\end{flalign} 
\end{subequations}
where $\Phi_{6} \triangleq \{ \eta_{m}, \mathbf{w}_{m}[n], \omega_{m}[n], \varpi_{m}[n], \tilde{\omega}_{m,j}[n],\tilde{\varpi}_{j}[n]\}$. 
Note that problem (\textbf{P6}) is still non-convex due to the non-convex constraints \eqref{p2e} and \eqref{p2f}. Then, we apply the first-order Taylor expansion to obtain the lower bound for $\left| \mathbf{\overline{h}}_{\tilde{m}}^{\mathsf{H}}[n]\mathbf{w}_{i}[n] \right|^{2}$ and $\| \mathbf{w}_{i}[n] \|^{2}$, i.e., 
\begin{align}
\left| \mathbf{\overline{h}}_{\tilde{m}}^{\mathsf{H}}[n] \mathbf{w}_{i}[n] \right|^{2} &\geq  \left| \mathbf{\overline{h}}_{\tilde{m}}^{\mathsf{H}}[n]\mathbf{w}_{i}^{(l)}[n] \right|^{2} + 2\Re \bigg\{ \left(\mathbf{w}_{i}^{(l)}[n]\right)^{\mathsf{H}} \nonumber \\
&\quad \times \mathbf{\overline{h}}_{\tilde{m}}[n] \mathbf{\overline{h}}_{\tilde{m}}^{\mathsf{H}}[n] \left(\mathbf{w}_{i}[n]-\mathbf{w}_{i}^{(l)}[n] \right) \bigg\} \nonumber \\
& \triangleq C_{\tilde{m},i}^{(l)}[n], 
\end{align}
\begin{align}
\|\mathbf{w}_i[n]\|^2 &\geq  \|\mathbf{w}_{i}^{(l)}[n]\|^2 \nonumber \\
&\quad + 2 \Re \left\{ \left(\mathbf{w}_{i}^{(l)}[n] \right)^{\mathsf{H}} \left( \mathbf{w}_{i}[n] - \mathbf{w}_{i}^{(l)}[n] \right) \right\} \nonumber \\
& \triangleq D_{i}^{(l)}[n],
\end{align}
where $\tilde{m} \in \mathcal{M} \bigcup \mathcal{J}$. 
Therefore, constraints \eqref{p2e} and \eqref{p2f} can be reformulated as 
\begin{align}
& \sum\limits_{i=1}^{\mathcal{M}} \left(  \zeta^{\rm{los}}_{m}[n]  C_{m,i}^{(l)}[n] + \zeta^{\rm{nlos}}_{m}[n] D_{i}^{(l)}[n] \right) + \sigma^2_{m} \nonumber \\
& \hspace{4em} \geq \frac{1}{2} \left( \frac{\omega_{m}^{(l)}[n]}{\varpi_{m}^{(l)}[n]} \varpi_{m}^{2}[n] + \frac{\varpi_{m}^{(l)}[n]}{\omega_{m}^{(l)}[n]} \omega_{m}^{2}[n] \right) , 
\label{p6b}
\end{align}
\begin{align}
&\sum\limits_{i=1,i\neq m }^{ \mathcal{M} } \left(  \zeta^{\rm{los}}_{j}[n]  C_{j,i}^{(l)}[n] + \zeta^{\rm{nlos}}_{j}[n] D_{i}^{(l)}[n] \right) + \sigma^2_{j} \nonumber \\
& \hspace{4em} \geq \frac{1}{2} \left( \frac{\tilde{\omega}_{m,j}^{(l)}[n]}{\tilde{\varpi}_{j}^{(l)}[n]} \tilde{\varpi}_{j}^{2}[n] + \frac{\tilde{\varpi}_{j}^{(l)}[n]}{\tilde{\omega}_{m,j}^{(l)}[n]} \tilde{\omega}_{m,j}^{2}[n] \right).
\label{p6c}
\end{align}

With the above transformations, problem (\textbf{P6}) can be reformulated as the following problem
\begin{subequations} 
\begin{flalign}
(\textbf{P7}):\ \max_{ \Phi_{7} } \, & \sum_{m=1}^{M}\sum_{n=1}^{N}  \tilde{R}_{m}^{\rm{lb}}[n] + \sum_{m=1}^{M} \eta_{m}   \\
{\rm{s.t.}}  \, &\eqref{p1d}, \eqref{p2c}, \eqref{p2g}-\eqref{p2h}, \eqref{p6b}-\eqref{p6c}, \nonumber
\end{flalign} 
\end{subequations}
where $\Phi_{7} \triangleq \{ \eta_{m}, \mathbf{w}_{m}[n], \omega_{m}[n], \varpi_{m}[n], \tilde{\omega}_{m,j}[n],\tilde{\varpi}_{j}[n]\}$.
Obviously, problem (\textbf{P7}) is a standard convex optimization problem, which can be efficiently solved via CVX \cite{grant2008cvx}.

\subsection{FA Positions Optimization}
Given $\{ \mathbf{q}[n], \mathbf{w}_{m}[n] \}$, the subproblem of (\textbf{P3}) for optimizing the FA positions $\{ x_{k}[n] \}$ can be expressed as
\begin{subequations} 
\begin{flalign}
(\textbf{P8}):\ \max_{ \Phi_{8} } \, & \sum_{m=1}^{M}\sum_{n=1}^{N}  \tilde{R}_{m}^{\rm{lb}}[n] + \sum_{m=1}^{M} \eta_{m}  \label{p8a}  \\
{\rm{s.t.}}  \, &\eqref{p1e}-\eqref{p1g}, \eqref{p2c}, \eqref{p2e}, \eqref{p2f}-\eqref{p2h}, \nonumber
\end{flalign} 
\end{subequations}
where $\Phi_{8} \triangleq \{ \eta_{m}, x_{k}[n], \omega_{m}[n], \varpi_{m}[n], \tilde{\omega}_{m,j}[n],\tilde{\varpi}_{j}[n]\}$. 
Note that problem (\textbf{P8}) is non-convex due to the non-convex constraints \eqref{p2c}, \eqref{p2e}, \eqref{p2f}, and \eqref{p2g}.

For convenience, constraint \eqref{p1g} can be equivalently expressed as
\begin{equation}
x_{k}[n-1] - t_{1}[n] \tilde{v} \leq x_{k}[n] \leq  x_{k}[n-1] + t_{1}[n] \tilde{v} . \label{p8b}
\end{equation}

Next, we analyze the upper and lower bounds of the term $\left| \mathbf{\overline{h}}_{\tilde{m}}^{\mathsf{H}}[n]\mathbf{w}_{i}[n] \right|^{2}, \tilde{m} \in \mathcal{M} \bigcup \mathcal{J}$ with respect to the FA positions $x_{k}[n]$.
Let $w_{i,k}[n] = |w_{i,k}[n]| e^{j \angle w_{i,k}[n]}$ denote the $k$-th element of the beamforming vector $\mathbf{w}_{i}[n]$ in time slot $n$, with amplitude $|w_{i,k}[n]|$ and phase $\angle w_{i,k}[n]$. Moreover, we define $\vartheta_{\tilde{m}}[n] = \frac{2\pi}{\lambda}\sin(\theta_{\tilde{m}}[n])$.
Then, by expanding the term $\left| \mathbf{\overline{h}}_{\tilde{m}}^{\mathsf{H}}[n]\mathbf{w}_{i}[n] \right|^{2}$, we can obtain 
\begin{align}
\left| \mathbf{\overline{h}}_{\tilde{m}}^{\mathsf{H}}[n]\mathbf{w}_{i}[n] \right|^{2}
&= \left| \sum_{p=1}^{K} w_{i,p}^{*}[n]  e^{{\mathrm{j}} \vartheta_{\tilde{m}}[n] x_{p}[n]  }  \right|^{2} 
\nonumber \\
& = \sum_{p=1}^{K} \sum_{q=1}^{K} 
\left| w_{i,p}[n] w_{i,q}[n] \right|
\cos \Lambda_{\tilde{m},i,p,q}[n],
\end{align} 
where $ \Lambda_{\tilde{m},i,p,q}[n] = \vartheta_{\tilde{m}}[n] (x_{p}[n]-x_{q}[n]) -( \angle {w}_{i,p}[n] - \angle {w}_{i,q}[n]) $.

Since $\left| \mathbf{\overline{h}}_{\tilde{m}}^{\mathsf{H}}[n]\mathbf{w}_{i}[n] \right|^{2}$ is non-concave and non-convex with respect to $x_{k}[n]$, we construct a lower bound surrogate function to locally approximate it using its second-order Taylor expansion.
Specifically, for a given $a \in \mathbb{R}$, we can obtain
\begin{align}
\cos (a) &\approx \cos (a_0) - \sin (a_0) (a -a_0) - \frac{1}{2} \cos (a_0) (a -a_0)^{2} \nonumber \\
& \geq \cos (a_0) - \sin (a_0) (a -a_0) - \frac{1}{2} (a -a_0)^{2} \nonumber \\
&= f\left(a \mid a_0 \right).
\end{align}
Given the local point $x_{k}^{(l)}[n]$, by letting $a_0 = \Lambda_{\tilde{m},i,p,q}^{(l)}[n]$ and $ a = \Lambda_{\tilde{m},i,p,q}[n]$, we can obtain
\begin{align}
&\left| \mathbf{\overline{h}}_{\tilde{m}}^{\mathsf{H}}[n]\mathbf{w}_{i}[n] \right|^{2} \nonumber \\
& \geq \sum_{p=1}^{K} \sum_{q=1}^{K} 
\left| w_{i,p}[n] w_{i,q}[n] \right|  
f\left(\Lambda_{\tilde{m},i,p,q}[n] \mid \Lambda_{\tilde{m},i,p,q}^{(l)}[n] \right) \nonumber \\
& \geq \frac{1}{2} \mathbf{x}^{\mathsf{T}}[n] \mathbf{S}^{\rm{lb}}_{\tilde{m},i}[n] \mathbf{x}[n] +\left(\mathbf{t}_{\tilde{m},i}^{\rm{lb}}[n]\right)^{\mathsf{T}} \mathbf{x}[n] + u^{\rm{lb}}_{\tilde{m},i}[n],
\end{align}
where $\mathbf{x}[n] = [x_{1}[n],\dots, x_{K}[n]]^{\mathsf{T}}$ denotes the FA position vector in time slot $n$.
$\mathbf{S}^{\rm{lb}}_{\tilde{m},i}[n] \in \mathbb{R}^{K \times K}$, $\mathbf{t}^{\rm{lb}}_{\tilde{m},i}[n] \in \mathbb{R}^{K \times 1}$, and $u^{\rm{lb}}_{\tilde{m},i}[n] \in \mathbb{R}$ are defined as 
\begin{align}
\mathbf{S}^{\rm{lb}}_{\tilde{m},i}[n] = -2 \vartheta_{\tilde{m}}^{2}[n]
\left( \mu_{i}[n] \mathsf{diag} \left(\mathbf{\tilde{w}}_{i}[n] \right) -\mathbf{\tilde{w}}_{i}[n] \mathbf{\tilde{w}}_{i}^{\mathsf{T}}[n] \right), 
\end{align}
\begin{align}
[\mathbf{t}^{\rm{lb}}_{\tilde{m},i}]_{p}[n] =&  2 \vartheta_{\tilde{m}}^{2}[n] \sum_{q=1}^{K} |w_{i,p}[n] w_{i,q}[n]| \left( x_{p}^{(l)}[n] - x_{q}^{(l)}[n] \right) \nonumber \\
& - 2 \vartheta_{\tilde{m}}[n] \sum_{q=1}^{K} |w_{i,p}[n] w_{i,q}[n]| \sin \left( \Lambda_{\tilde{m},i,p,q}^{(l)}[n]  \right),
\end{align} 
\begin{align}
&u^{\rm{lb}}_{\tilde{m},i}[n] \nonumber \\
&= \sum_{p=1}^{K} \sum_{q=1}^{K} |w_{i,p}[n] w_{i,q}[n]|  \cos \left( \Lambda_{\tilde{m},i,p,q}^{(l)}[n] \right) \nonumber \\
& \quad + \vartheta_{\tilde{m}}[n] \sum_{p=1}^{K} \sum_{q=1}^{K} |w_{i,p}[n] w_{i,q}[n]| \sin \left( \Lambda_{\tilde{m},i,p,q}^{(l)}[n] \right) \nonumber \\
&\quad \times \left( x_{p}^{(l)}[n] - x_{q}^{(l)}[n] \right) \nonumber \\
&\quad - \frac{1}{2} \vartheta_{\tilde{m}}^{2}[n]
\sum_{p=1}^{K} \sum_{q=1}^{K} |w_{i,p}[n] w_{i,q}[n]|
\left(x_{p}^{(l)}[n] - x_{q}^{(l)}[n]\right)^{2} ,
\end{align}
where $\mathbf{\tilde{w}}_{i}[n] = \left[|w_{i,1}[n]|,|w_{i,2}[n]|,\cdots, |w_{i,K}[n]| \right]^{\mathsf{T}}$, and $\mu_{i}[n]$
$ = \sum_{p = 1}^{K} |w_{i,p}[n]|$. By introducing the slack variable $\phi_{\tilde{m},i}^{\rm{lb}}[n]$, we can obtain 
\begin{equation} 
\frac{1}{2} \mathbf{x}^{\mathsf{T}}[n] \mathbf{S}^{\rm{lb}}_{\tilde{m},i}[n] \mathbf{x}[n] +\left(\mathbf{t}_{\tilde{m},i}^{\rm{lb}}[n]\right)^{\mathsf{T}} \mathbf{x}[n] + u^{\rm{lb}}_{\tilde{m},i}[n] \geq \phi_{\tilde{m},i}^{\rm{lb}}[n]. \label{p8c}
\end{equation}
Since $\mathbf{S}^{\rm{lb}}_{\tilde{m},i}[n]$ is negative semi-definite, constraint \eqref{p8c} is convex with respect to $\mathbf{x}[n]$. The complete proof is provided in Appendix \ref{proof2}.

Similarly, we can get an upper bound surrogate function to locally approximate $\left| \mathbf{\overline{h}}_{\tilde{m}}^{\mathsf{H}}[n]\mathbf{w}_{i}[n] \right|^{2}$ using its second-order Taylor expansion.
Specifically, for a given $a \in \mathbb{R}$, we can obtain
\begin{align}
\cos (a) &\approx \cos (a_0) - \sin (a_0) (a -a_0) - \frac{1}{2} \cos (a_0) (a -a_0)^{2} \nonumber \\
& \leq \cos (a_0) - \sin (a_0) (a -a_0) + \frac{1}{2} (a -a_0)^{2} \nonumber \\
&= \tilde{f}\left(a \mid a_0 \right).
\end{align}
By letting $a_0 = \Lambda_{\tilde{m},i,p,q}^{(l)}[n]$ and $ a = \Lambda_{\tilde{m},i,p,q}[n]$, we can obtain
\begin{align}
&\left| \mathbf{\overline{h}}_{\tilde{m}}^{\mathsf{H}}[n]\mathbf{w}_{i}[n] \right|^{2} \nonumber \\
& \leq \sum_{p=1}^{K} \sum_{q=1}^{K} 
\left| w_{i,p}[n] w_{i,q}[n] \right|  
\tilde{f} \left(\Lambda_{\tilde{m},i,p,q}[n] \mid \Lambda_{\tilde{m},i,p,q}^{(l)}[n] \right) \nonumber \\
& \leq \frac{1}{2} \mathbf{x}^{\mathsf{T}}[n] \mathbf{S}^{\rm{ub}}_{\tilde{m},i}[n] \mathbf{x}[n] +\left(\mathbf{t}_{\tilde{m},i}^{\rm{ub}}[n]\right)^{\mathsf{T}} \mathbf{x}[n] + u^{\rm{ub}}_{\tilde{m},i}[n],
\end{align}
where $\mathbf{S}^{\rm{ub}}_{\tilde{m},i}[n] \in \mathbb{R}^{K \times K}$, $\mathbf{t}^{\rm{ub}}_{\tilde{m},I}[n] \in \mathbb{R}^{K \times 1}$, and $u^{\rm{ub}}_{\tilde{m},i}[n] \in \mathbb{R}$ are defined as 
\begin{align}
\mathbf{S}^{\rm{ub}}_{\tilde{m},i}[n] = 2 \vartheta_{\tilde{m}}^{2}[n]
\left( \mu_{i}[n] \mathsf{diag} \left(\mathbf{\tilde{w}}_{i}[n] \right) -\mathbf{\tilde{w}}_{i}[n] \mathbf{\tilde{w}}_{i}^{\mathsf{T}}[n] \right), 
\end{align}
\begin{align}
[\mathbf{t}^{\rm{ub}}_{\tilde{m},i}]_{p}[n] =& -2 \vartheta_{\tilde{m}}^{2}[n] \sum_{q=1}^{K} |w_{i,p}[n] w_{i,q}[n]| \left( x_{p}^{(l)}[n] - x_{q}^{(l)}[n] \right) \nonumber \\
& - 2 \vartheta_{\tilde{m}}[n] \sum_{q=1}^{K} |w_{i,p}[n] w_{i,q}[n]| \sin \left( \Lambda_{\tilde{m},i,p,q}^{(l)}[n]  \right),
\end{align} 
\begin{align}
&u^{\rm{ub}}_{\tilde{m},i}[n] \nonumber \\
&= \sum_{p=1}^{K} \sum_{q=1}^{K} |w_{i,p}[n] w_{i,q}[n]|  \cos \left( \Lambda_{\tilde{m},i,p,q}^{(l)}[n] \right) \nonumber \\
& \quad + \vartheta_{\tilde{m}}[n] \sum_{p=1}^{K} \sum_{q=1}^{K} |w_{i,p}[n] w_{i,q}[n]| \sin \left( \Lambda_{\tilde{m},i,p,q}^{(l)}[n] \right) \nonumber  \\
&\quad \times \left( x_{p}^{(l)}[n] - x_{q}^{(l)}[n] \right) \nonumber \\
&\quad + \frac{1}{2} \vartheta_{\tilde{m}}^{2}[n]
\sum_{p=1}^{K} \sum_{q=1}^{K} |w_{i,p}[n] w_{i,q}[n]|
\left(x_{p}^{(l)}[n] - x_{q}^{(l)}[n]\right)^{2} .
\end{align}
By introducing the slack variable $\phi_{\tilde{m},i}^{\rm{ub}}[n]$, we can obtain 
\begin{equation} 
\frac{1}{2} \mathbf{x}^{\mathsf{T}}[n] \mathbf{S}^{\rm{ub}}_{\tilde{m},i}[n] \mathbf{x}[n] +\left(\mathbf{t}_{\tilde{m},i}^{\rm{ub}}[n]\right)^{\mathsf{T}} \mathbf{x}[n] + u^{\rm{ub}}_{\tilde{m},i}[n] \leq \phi_{\tilde{m},i}^{\rm{ub}}[n]. \label{p8d}
\end{equation}
Since $\mathbf{S}^{\rm{ub}}_{\tilde{m},i}[n]$ is positive semi-definite, constraint \eqref{p8d} is convex with respect to $\mathbf{x}[n]$.

Therefore, constraints \eqref{p2c}, \eqref{p2e}, \eqref{p2f}, and \eqref{p2g} can be reformulated as 
\begin{equation}
\varpi_{m}[n] \geq \sum\limits_{ i = 1, i \neq m }^{ \mathcal{M} } \left( \zeta^{\rm{los}}_{m}[n] \phi_{m,i}^{\rm{ub}}[n] + \zeta^{\rm{nlos}}_{m}[n] \| \mathbf{w}_{i}[n] \|^{2} \right) + \sigma^2_{m},
\label{p8e}
\end{equation}
\begin{align}
&\sum\limits_{ i = 1 }^{ \mathcal{M} } \left( \zeta^{\rm{los}}_{m}[n] \phi_{m,i}^{\rm{lb}}[n] + \zeta^{\rm{nlos}}_{m}[n] \| \mathbf{w}_{i}[n] \|^{2} \right) + \sigma^2_{m} \nonumber \\
&\hspace{4em}  \geq \frac{1}{2} \left( \frac{\omega_{m}^{(l)}[n]}{\varpi_{m}^{(l)}[n]} \varpi_{m}^{2}[n] + \frac{\varpi_{m}^{(l)}[n]}{\omega_{m}^{(l)}[n]} \omega_{m}^{2}[n] \right), 
\label{p8f}
\end{align}
\begin{align}
&\sum\limits_{ i = 1,i \neq m }^{ \mathcal{M} } \left( \zeta^{\rm{los}}_{j}[n] \phi_{j,i}^{\rm{lb}}[n] + \zeta^{\rm{nlos}}_{j}[n] \| \mathbf{w}_{i}[n] \|^{2} \right) + \sigma^2_{j} \nonumber \\
&\hspace{4em} \geq \frac{1}{2} \left( \frac{\tilde{\omega}_{m,j}^{(l)}[n]}{\tilde{\varpi}_{j}^{(l)}[n]} \tilde{\varpi}_{j}^{2}[n] + \frac{\tilde{\varpi}_{j}^{(l)}[n]}{\tilde{\omega}_{m,j}^{(l)}[n]} \tilde{\omega}_{m,j}^{2}[n] \right),
\label{p8g}
\end{align}
\begin{equation}
\tilde{\varpi}_{j}[n] \geq \sum\limits_{ i = 1 }^{ \mathcal{M} } \left( \zeta^{\rm{los}}_{j}[n] \phi_{j,i}^{\rm{ub}}[n] + \zeta^{\rm{nlos}}_{j}[n] \| \mathbf{w}_{i}[n] \|^{2} \right) + \sigma^2_{j}.
\label{p8h}
\end{equation}

With the above transformations, problem (\textbf{P8}) can be reformulated as the following problem 
\begin{subequations} 
\begin{flalign}
(\textbf{P9}):\ \max_{ \Phi_{9} } \, & \sum_{m=1}^{M}\sum_{n=1}^{N}  \tilde{R}_{m}^{\rm{lb}}[n] + \sum_{m=1}^{M} \eta_{m}  \label{p8a}  \\
{\rm{s.t.}}  \, &\eqref{p1e}-\eqref{p1f}, \eqref{p2h}, \eqref{p8b}, \eqref{p8c}, \eqref{p8d}-\eqref{p8h}, \nonumber
\end{flalign} 
\end{subequations}
where $\Phi_{9} \triangleq \{ \eta_{m}, x_{k}[n], \omega_{m}[n], \varpi_{m}[n], \tilde{\omega}_{m,j}[n],\tilde{\varpi}_{j}[n], \phi_{\tilde{m},i}^{\rm{lb}}[n], $
$\phi_{\tilde{m},i}^{\rm{ub}}[n]\}$. 
Obviously, problem (\textbf{P9}) is a standard convex optimization problem, which can be efficiently solved via CVX \cite{grant2008cvx}.

\subsection{Initialization, Convergence and Complexity Analysis}
Based on the above derivations, the proposed AO algorithm for the FA-enabled secure UAV communication system is summarized in Algorithm \ref{algorithm1}. 

\begin{algorithm}[t] \small
\caption{AO Algorithm for Solving Problem (\textbf{P1}).} 
\label{algorithm1} 
\begin{algorithmic}[1]
\STATE
\textbf{Initialize:} $\{ \mathbf{q}^{(0)}[n],\mathbf{w}_{m}^{(0)}[n],x_{k}^{(0)}[n] \}$, iteration index $ l = 1 $, maximum iteration number $l_{\rm{max}}$, accuracy threshold $ \varepsilon > 0 $. 
\STATE \textbf{Repeat:}
\STATE With obtained $\{\mathbf{w}_{m}^{(l-1)}[n]\}$, $\{x_{k}^{(l-1)}[n]\}$, and auxiliary variables, 
solve the problem (\textbf{P5}) to obtain $\{\mathbf{q}^{(l)}[n]\}$.
\STATE With obtained $\{\mathbf{q}^{(l)}[n]\}$, $\{x_{k}^{(l-1)}[n]\}$, and auxiliary variables, 
solve the problem (\textbf{P7}) to obtain $\{\mathbf{w}_{m}^{(l)}[n]\}$.
\STATE With obtained $\{\mathbf{q}^{(l)}[n]\}$, $\{\mathbf{w}_{m}^{(l)}[n]\}$, and auxiliary variables, 
solve the problem (\textbf{P9}) to obtain $\{x_{k}^{(l)}[n]\}$.
\STATE According to $\{\mathbf{q}^{(l)}[n], \mathbf{w}_{m}^{(l)}[n], x_{k}^{(l)}[n]\}$, update the objective value $\mathcal{F}\left( \mathbf{q}^{(l)}[n], \mathbf{w}_{m}^{(l)}[n], x_{k}^{(l)}[n] \right)$.
\STATE Update $ l = l + 1 $.
\STATE
\textbf{Until:} the increase of the value of the objective between two adjacent iterations is smaller than $ \varepsilon $ or $l > l_{\max}$.
\STATE \textbf{Output:} $ \mathbf{q}^{(*)}[n],\mathbf{w}_{m}^{(*)}[n],x_{k}^{(*)}[n]$
\end{algorithmic} 
\end{algorithm}

\subsubsection{Initialization Analysis} 
For AO-based algorithms, the initialization of variables plays a critical role in determining convergence speed and solution quality. In this paper, the UAV trajectory is initialized as a smooth curve that remains in proximity to the users, as illustrated in Fig. \ref{UAV trajectory} for the trajectory of the FT I scheme. The FA positions are uniformly distributed over the interval $[0,D_{\rm{FA}}]$ with spacing $\frac{D_{\rm{FA}}}{K-1}$. The transmit beamforming is initialized following the maximum ratio transmission (MRT) principle, where the initial beamforming directions are determined based on the LoS channel components between the UAV and the users.

\subsubsection{Initialization of Auxiliary Variables}
Given the initial UAV trajectory, transmit beamforming, and FA positions, the auxiliary variables are initialized based on the corresponding statistical received signal and interference powers. Specifically, according to \eqref{p2b} and \eqref{p2c}, the initial values of $\varpi_m[n]$ and $\omega_m[n]$ are respectively given by
\begin{equation}\small
\varpi_{m}^{(0)}[n]=\sum_{\substack{i=1,i\neq m}}^{M}
A_{m,i}^{(0)}[n]+\sigma_{m}^{2}, \,\,
\omega_{m}^{(0)}[n]=\frac{\sum_{i=1}^{M}A_{m,i}^{(0)}[n]+\sigma_{m}^{2}
}{\varpi_{m}^{(0)}[n]}
\end{equation}
Similarly, the auxiliary variables associated with the sampled eavesdropper locations are initialized as
\begin{equation}\small
\tilde{\varpi}_{j}[n]=\sum_{i=1}^{M}B_{j,i}^{(0)}[n]+\sigma_{j}^{2}, \,\,
\tilde{\omega}_{m,j}[n] =\frac{\sum_{\substack{i=1,i\neq m}}^{M}
B_{j,i}^{(0)}[n]+\sigma_{j}^{2}}{\tilde{\varpi}_{j}[n]}.
\end{equation}
Here, $A_{m,i}^{(0)}[n]$ and $B_{j,i}^{(0)}[n]$ are evaluated using the initial optimization variables. These initial values satisfy the corresponding auxiliary-variable constraints with equality and are updated based on the current optimization variables after each AO update. The remaining slack variables are directly determined within their corresponding convex optimization subproblems and thus do not require separate initialization.

\subsubsection{Convergence Analysis} 
For any optimization algorithm, analyzing the convergence of the algorithm is essential. The objective is defined as $\mathcal{F}\left( \mathbf{q}[n], \mathbf{w}_{m}[n], x_{k}[n] \right)$. 
In step 3 of Algorithm \ref{algorithm1}, $\{\mathbf{q}^{(l)}[n] \}$ can be obtained. Thus, we have
\begin{align}
& \mathcal{F}\left( \mathbf{q}^{(l-1)}[n], \mathbf{w}_{m}^{(l-1)}[n], x_{k}^{(l-1)}[n] \right) \nonumber \\
\leq& \mathcal{F}\left( \mathbf{q}^{(l)}[n], \mathbf{w}_{m}^{(l-1)}[n], x_{k}^{(l-1)}[n] \right) .
\end{align}
Similarly, from the steps 4-5 of Algorithm \ref{algorithm1}, we can obtain
\begin{align}
& \mathcal{F}\left( \mathbf{q}^{(l)}[n], \mathbf{w}_{m}^{(l-1)}[n], x_{k}^{(l-1)}[n] \right)  \nonumber \\
\leq& \mathcal{F}\left( \mathbf{q}^{(l)}[n], \mathbf{w}_{m}^{(l)}[n], x_{k}^{(l-1)}[n] \right)  \nonumber \\
\leq& \mathcal{F}\left( \mathbf{q}^{(l)}[n], \mathbf{w}_{m}^{(l)}[n], x_{k}^{(l)}[n] \right) ,
\end{align}
which demonstrates that the objective value is non-decreasing over iterations. Moreover, the objective value is upper-bounded due to the limited transmit power. Therefore, Algorithm \ref{algorithm1} is guaranteed to converge.

\subsubsection{Complexity Analysis} 
In Algorithm \ref{algorithm1}, subproblems (\textbf{P5}), (\textbf{P7}), and (\textbf{P9}) are solved using the interior point method \cite{10100680}. 
The computational complexities of these subproblems are $\mathcal{O}\big( N^{3.5}(M+J)^{3.5} \log(\varepsilon^{-1})\big)$, 
$\mathcal{O}\big( (MN)^{3.5}K^{3} \log(\varepsilon^{-1})\big)$, 
and $\mathcal{O}\big((NKM)^{3.5} \log(\varepsilon^{-1})\big)$, respectively. Accordingly, the overall computational complexity of Algorithm \ref{algorithm1} can be given by $\mathcal{O}( l_{\rm{max}} (NKM)^{3.5}\log(1/\epsilon) )$.

\begin{table}[t]
\centering
\caption{Simulation Parameters.}
\label{tab:parameters}
\begin{tabular}{p{6.2cm}c}
\toprule
\centering
Parameters & Value\\
\midrule  
\centering Mission period, $T$ &  45 s\\
\centering Number of time slots, $N$ & 20\\
\centering Duration of time slot, $\tau$ & 2.25 s\\
\centering Number of transmit antennas at UAV, $K$ & 6\\
\centering Channel power gain at reference distance of $1 \mathrm{m}$, $h_{0}$& -60 dB\\
\centering Noise power at user receivers, $\sigma_{m}^{2}$& -80 dBm\\
\centering Noise power at eavesdropper, $\sigma_{j}^{2}$& -80 dBm\\
\centering Maximum UAV transmit power, $P_{\rm{max}}$ & 30 dBm \\
\centering Antenna mobility region, $[0, D_{\rm{FA}}]$ & $ [0, 20 \lambda ]$ \\
\centering Carrier wavelength, $\lambda$ & 0.0107 $\mathrm{m}$\\
\centering Minimum inter-antenna distance, $d_{\rm{min}}$ & $0.5\lambda$ \\
\centering Duration of FA movement, $t_{1}[n]$ & 0.2 s \\
\centering FA movement speed, $\tilde{v}$ & 0.1 m/s \\
\centering Maximum UAV speed, $V_{\rm{max}}$& $20 \ {\mathrm{m/s}}$ \\
\centering UAV flight altitude, $H$ & 100 $\mathrm{m}$ \\
\centering Coefficients $c_{1}$, $c_{2}$ & 1, 4.3976 \\
\bottomrule
\end{tabular}
\end{table}

\section{Numerical Results}
\begin{figure}[t]
\centering
\includegraphics[width=0.95\linewidth]{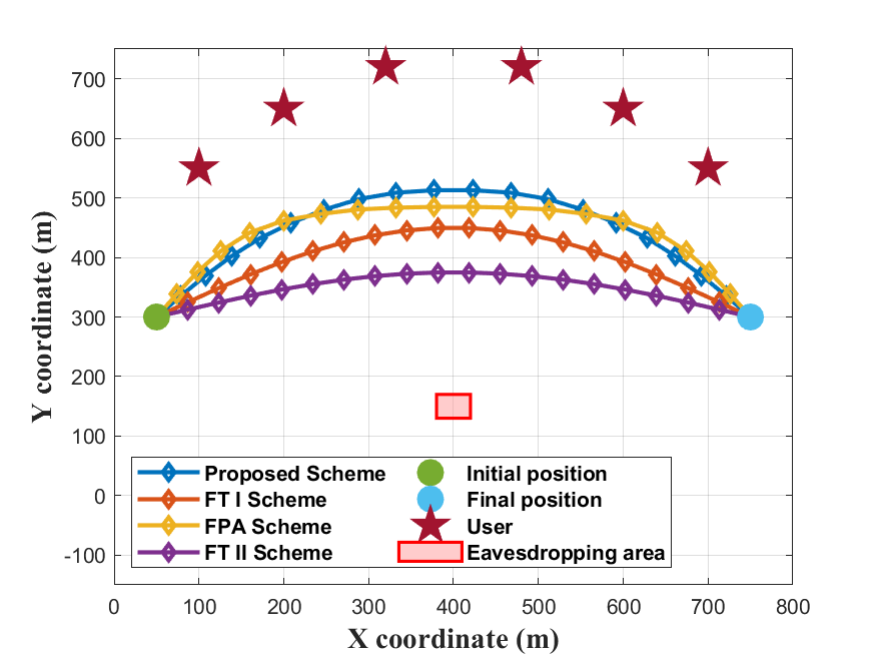}
\caption{UAV trajectories under different schemes.}
\label{UAV trajectory}
\end{figure}

This section presents numerical results to evaluate the performance of the proposed FA-enabled secure UAV communication system in terms of accumulated secrecy throughput. As shown in Fig.~\ref{UAV trajectory}, we consider an $800~\rm{m} \times 800~\rm{m}$ square region containing $M=6$ users and $J=10$ sampling locations within the eavesdropping region of $40~\rm{m} \times 40~\rm{m}$. The candidate eavesdropper locations are generated through independent uniform random sampling. To evaluate the performance gain, we compare the proposed scheme with the following five benchmark schemes:
\begin{itemize}
\item \textbf{PSO Scheme}: The FA positions are optimized using the particle swarm optimization (PSO) algorithm, while the UAV trajectory and transmit beamforming are optimized according to the proposed method. To limit the computational cost, the PSO algorithm is configured with a population size of 50 and a maximum of 50 iterations.
\item \textbf{RPA Scheme}: The FA positions are randomly generated in each time slot subject to the antenna spacing and movement constraints, while the UAV trajectory and transmit beamforming are jointly optimized using the proposed algorithm.
\item \textbf{FPA Scheme}: The FAs are fixed in a conventional half-wavelength spaced linear array configuration, while the UAV trajectory and transmit beamforming are optimized as per the proposed approach.
\item \textbf{FT I and FT II Schemes}: The UAV follows two different predefined trajectories, as illustrated in Fig. \ref{UAV trajectory}, while the FA positions and transmit beamforming are optimized using the proposed method.
\end{itemize}
Unless otherwise stated, the default simulation parameters are summarized in Table \ref{tab:parameters} \cite{11214460,10100680,11082461}.

\begin{figure}[t]
\centering
\includegraphics[width=0.95\linewidth]{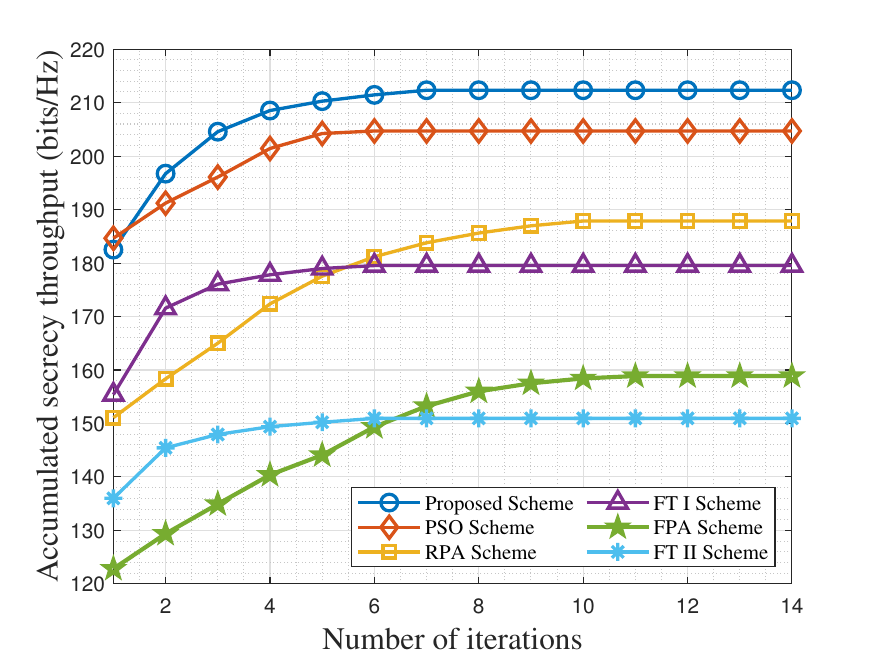}
\caption{The convergence performance of different schemes in terms of accumulated secrecy throughput.}
\label{convergence}
\end{figure}

\begin{figure*}[t]    
\centering            
\begin{subfigure}{0.32\textwidth}
\includegraphics[width=\linewidth]{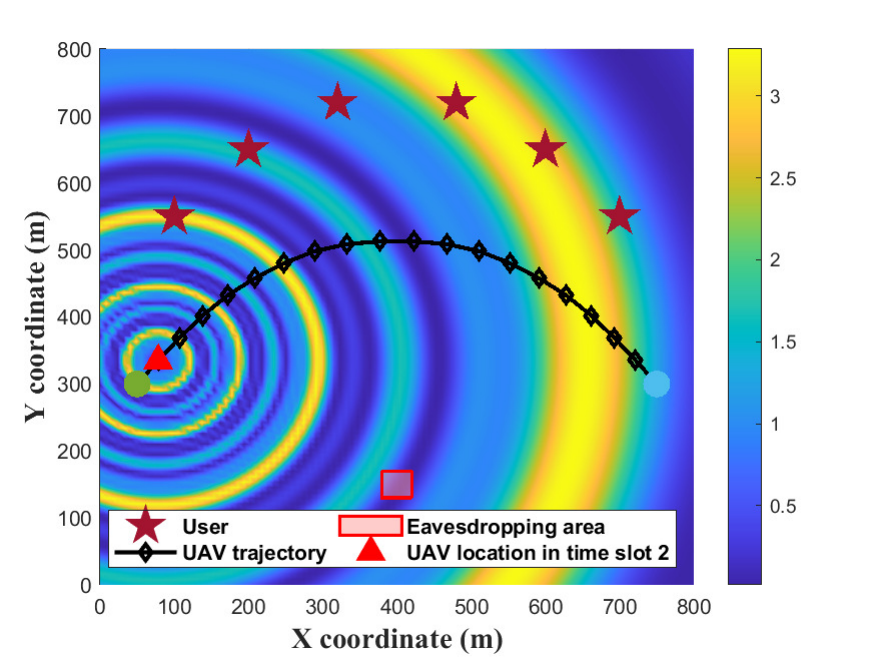}
\caption{$n$ = 2.} 
\end{subfigure}
\hfill
\begin{subfigure}{0.32\textwidth}
\includegraphics[width=\linewidth]{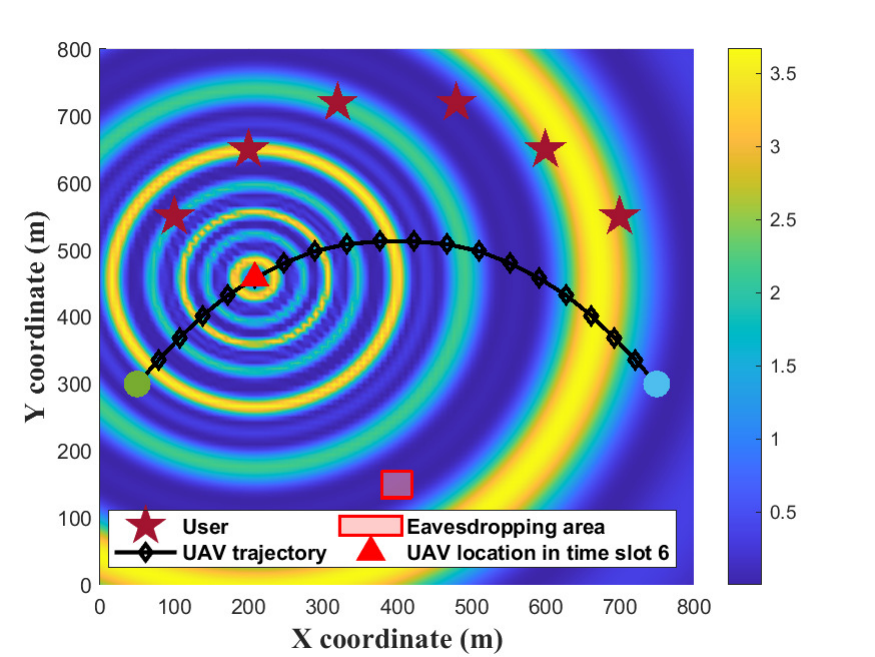}
\caption{$n$ = 6.}
\end{subfigure}
\hfill
\begin{subfigure}{0.32\textwidth}
\includegraphics[width=\linewidth]{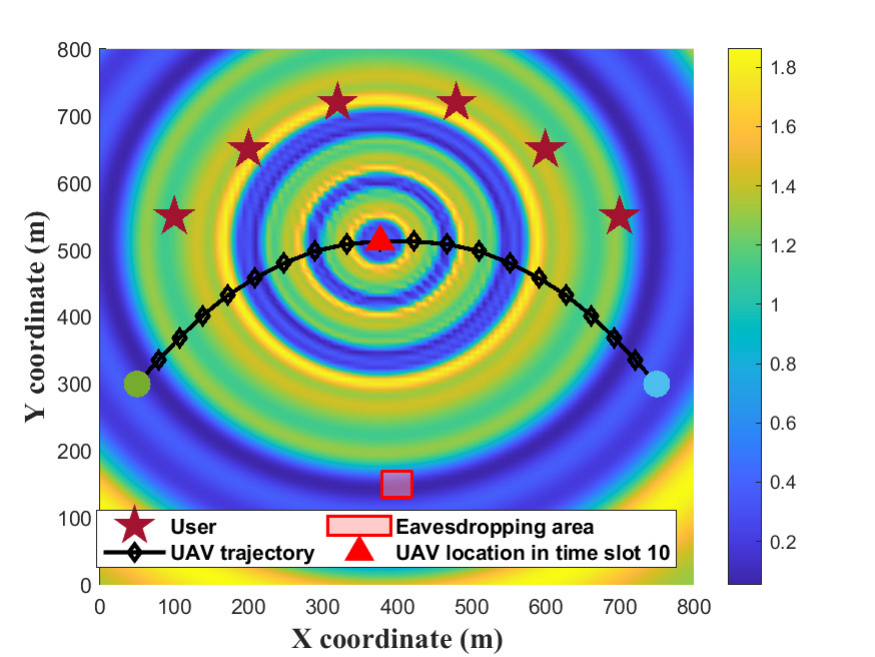}
\caption{$n$ = 10.}
\end{subfigure}
\caption{Beamforming gain patterns under the proposed scheme at the selected time slots $n$ = 2, 6, 10.}    
\label{fig:beamscheme1}            
\end{figure*}

\begin{figure*}[t]    
\centering            
\begin{subfigure}{0.32\textwidth}
\includegraphics[width=\linewidth]{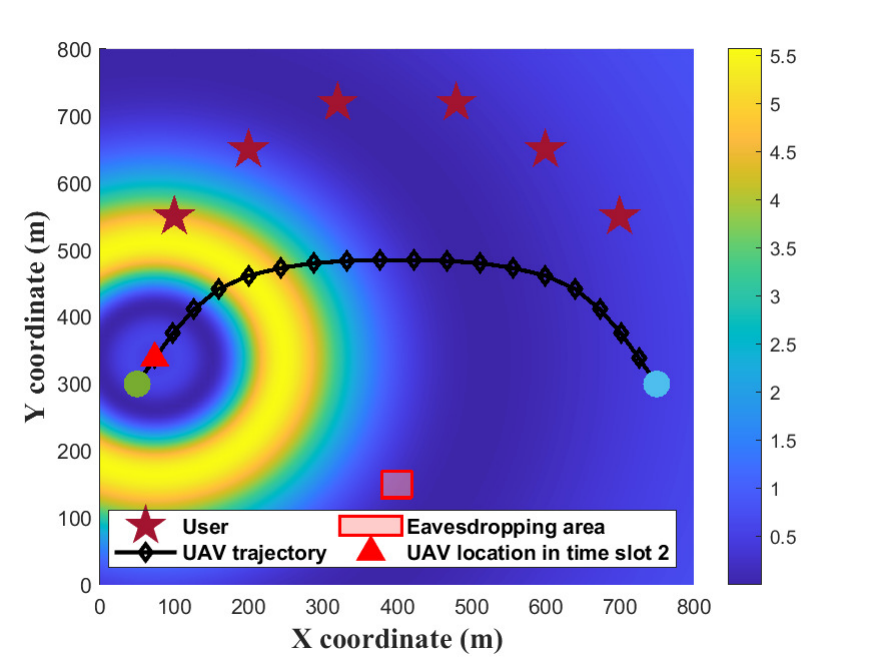}
\caption{$n$ = 2.} 
\end{subfigure}
\hfill
\begin{subfigure}{0.32\textwidth}
\includegraphics[width=\linewidth]{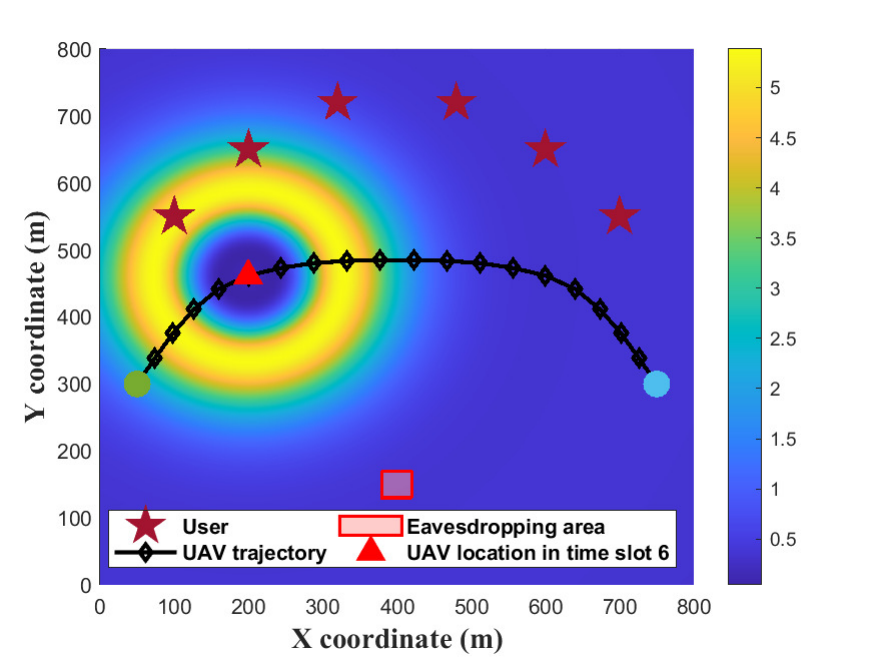}
\caption{$n$ = 6.}
\end{subfigure}
\hfill
\begin{subfigure}{0.32\textwidth}
\includegraphics[width=\linewidth]{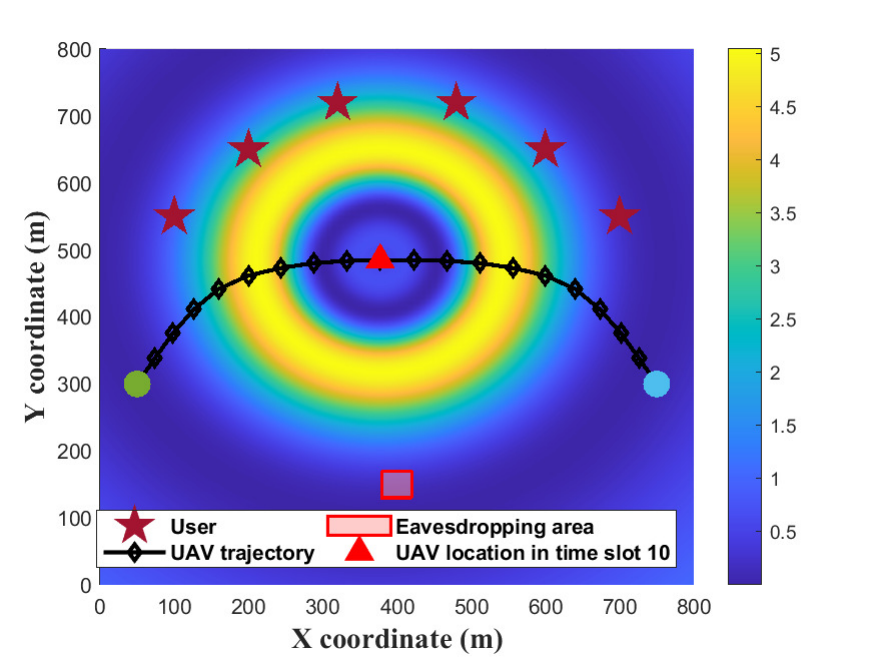}
\caption{$n$ = 10.}
\end{subfigure}
\caption{Beamforming gain patterns under the FPA scheme at the selected time slots $n$ = 2, 6, 10.}    
\label{fig:beamscheme2}            
\end{figure*}

\begin{figure*}[t]    
\centering            
\begin{subfigure}{0.32\textwidth}
\includegraphics[width=\linewidth]{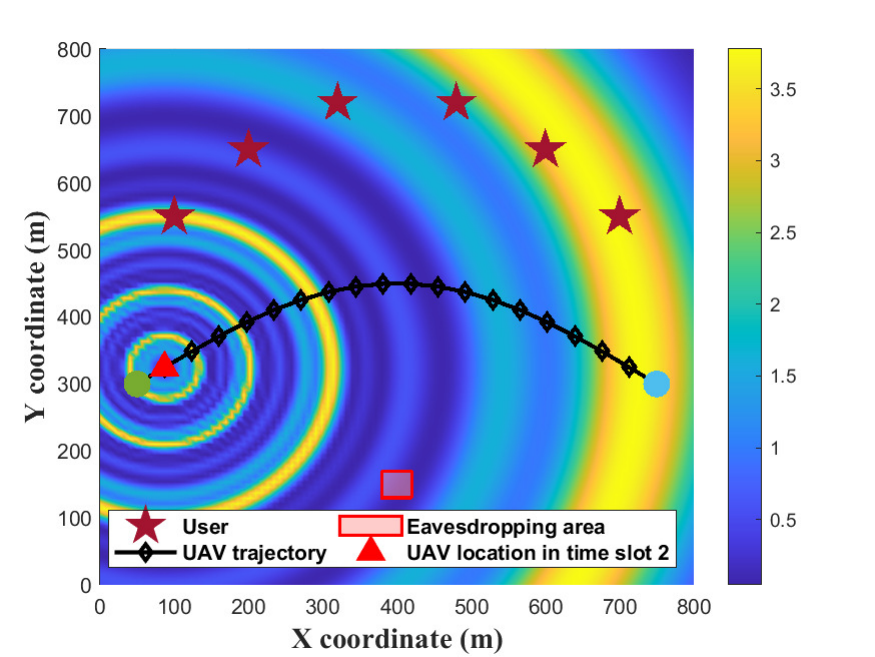}
\caption{$n$ = 2.} 
\end{subfigure}
\hfill
\begin{subfigure}{0.32\textwidth}
\includegraphics[width=\linewidth]{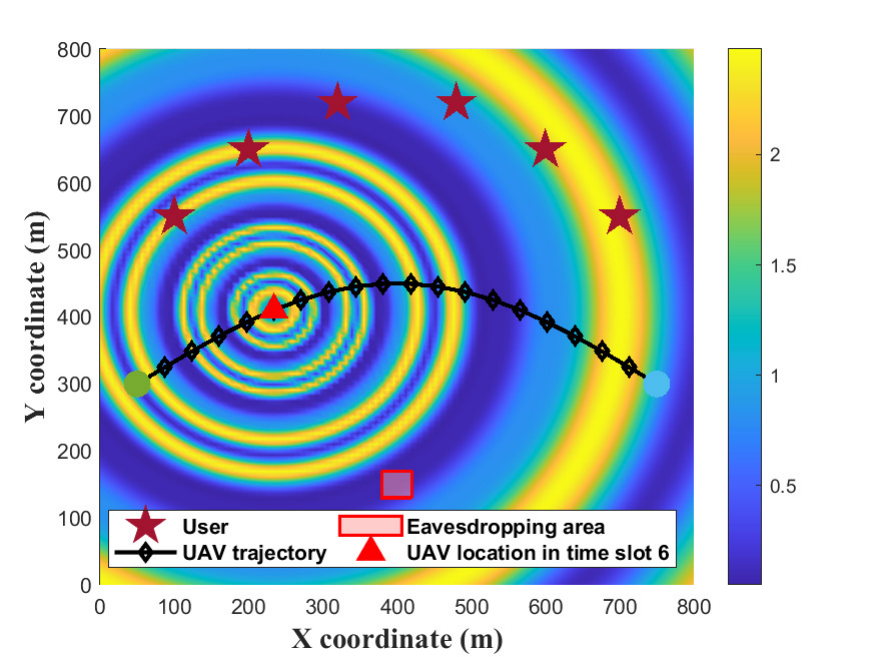}
\caption{$n$ = 6.}
\end{subfigure}
\hfill
\begin{subfigure}{0.32\textwidth}
\includegraphics[width=\linewidth]{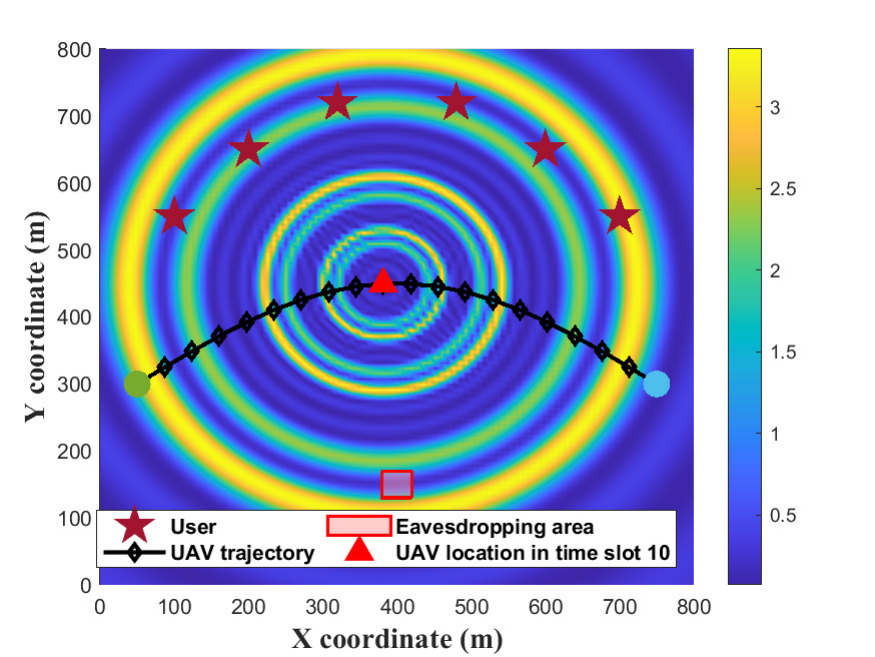}
\caption{$n$ = 10.}
\end{subfigure}
\caption{Beamforming gain patterns under the FT I scheme at the selected time slots $n$ = 2, 6, 10.}    
\label{fig:beamscheme3}            
\end{figure*}

Fig. \ref{convergence} illustrates the convergence behavior of the proposed algorithm and the benchmark schemes. The proposed scheme converges within approximately 8 iterations and achieves the highest accumulated secrecy throughput of about 212 bits/Hz, demonstrating its effectiveness. In comparison, the PSO scheme achieves a slightly lower accumulated secrecy throughput, with a performance loss of approximately 3.59\% under the considered computational budget. The RPA and FPA schemes exhibit lower performance, highlighting the benefits of optimizing FA positions. Furthermore, the FT I scheme outperforms the FPA scheme despite employing a predefined UAV trajectory, indicating that FA position reconfiguration can partially compensate for constrained UAV mobility. However, the lower performance of the FT II scheme suggests that the choice of UAV trajectory also plays an important role in the overall secrecy throughput.

Fig. \ref{UAV trajectory} illustrates the UAV trajectories under different schemes, together with the initial and final UAV positions, legitimate users, and the potential eavesdropping region. The proposed scheme and the FPA scheme follow smooth trajectories that remain relatively close to the legitimate users. Notably, although the predefined trajectory of the FT I scheme is farther from the users than that of the FPA scheme, the FT I scheme achieves higher accumulated secrecy throughput than that of the FPA scheme, as shown in Fig. \ref{convergence}. This observation indicates that FA position reconfiguration can partially compensate for a suboptimal UAV trajectory, highlighting the complementary benefits of UAV mobility and FA position optimization.

Figs. \ref{fig:beamscheme1}-\ref{fig:beamscheme3} illustrate the beamforming pattern gain achieved at selected time slots $n$ = 2, 6, 10 under three different schemes: the proposed scheme, the FPA scheme, and the FT I scheme. In Fig. \ref{fig:beamscheme1}, which corresponds to the proposed scheme, both the UAV trajectory and FA positions are jointly optimized with beamforming, enabling highly adaptive radiation patterns that dynamically concentrate energy toward the legitimate users (marked by red stars) while actively suppressing signal leakage into the uncertain eavesdropping region (highlighted by a red rectangle). As the UAV moves along its optimized trajectory, it travels from the left side at $n$ = 2 to a position near the user at $n$ = 6, and then maneuvers to maintain coverage at $n$ = 10, the beamforming pattern evolves accordingly, forming an increasingly narrow main lobe aligned with the users and creating deep nulls or low-gain zones over the eavesdropper area. 

Fig. \ref{fig:beamscheme2} illustrates the beamforming pattern gain of the FPA scheme. Despite UAV trajectory optimization, the half-wavelength spaced linear array configuration severely limits the system’s ability to sculpt the radiation pattern: the beam remains relatively broad across all time slots, with elevated sidelobes and noticeable power spillage into the eavesdropping region. Even when the UAV is well-positioned (e.g., at $n$ = 6), the lack of antenna element mobility prevents the formation of sharp directional beams or effective spatial nulls, resulting in inherently lower secrecy performance. This demonstrates that the FPA scheme, as in conventional phased arrays, leads to lower secure communication performance without spatial reconfigurability of the antenna array.

Fig. \ref{fig:beamscheme3} presents the beamforming pattern gain of the FT I scheme. Here, the dynamic reconfiguration of FAs partially compensates for the suboptimal UAV trajectory: the beamforming patterns exhibit improved focus toward the users and better suppression in the eavesdropping zone compared to the FPA scheme. At $n$ = 10, for instance, a discernible null emerges near the eavesdropper region, demonstrating the benefit of FAs. However, because the UAV trajectory is fixed, the system cannot exploit favorable aerial vantage locations or avoid geometries that inherently increase eavesdropping risk. Consequently, the beamforming patterns in Fig. \ref{fig:beamscheme3}, while superior to those in Fig. \ref{fig:beamscheme2}, still fall short of the fully adaptive strategy in Fig. \ref{fig:beamscheme1}, where the UAV actively navigates to positions that synergize with FA reconfiguration to maximize secrecy performance.

\begin{figure}[t]
\centering
\includegraphics[width=0.95\linewidth]{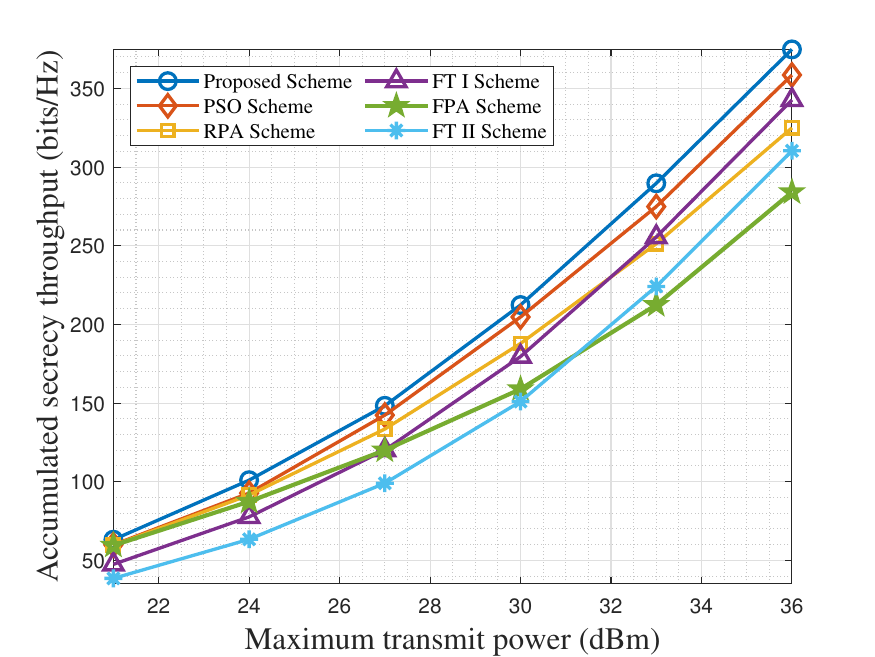}
\caption{Accumulated secrecy throughput versus the maximum transmit power.}
\label{transmit power}
\end{figure}

Fig. \ref{transmit power} illustrates the accumulated secrecy throughput of different schemes versus the maximum UAV transmit power, ranging from 21 dBm to 36 dBm. It can be observed that the accumulated secrecy throughput of all schemes increases with the transmit power, as higher transmit power generally improves the achievable rates of legitimate users. Moreover, the proposed scheme consistently outperforms the benchmark schemes over the considered power range, demonstrating the benefits of joint optimization. Interestingly, at relatively high transmit power levels, the FT II scheme achieves higher accumulated secrecy throughput than the FPA scheme, despite using a predefined UAV trajectory. This result suggests that FA position reconfiguration, together with transmit beamforming, can partially compensate for a suboptimal UAV trajectory, particularly when sufficient transmit power is available.

\begin{figure}[t]
\centering
\includegraphics[width=0.95\linewidth]{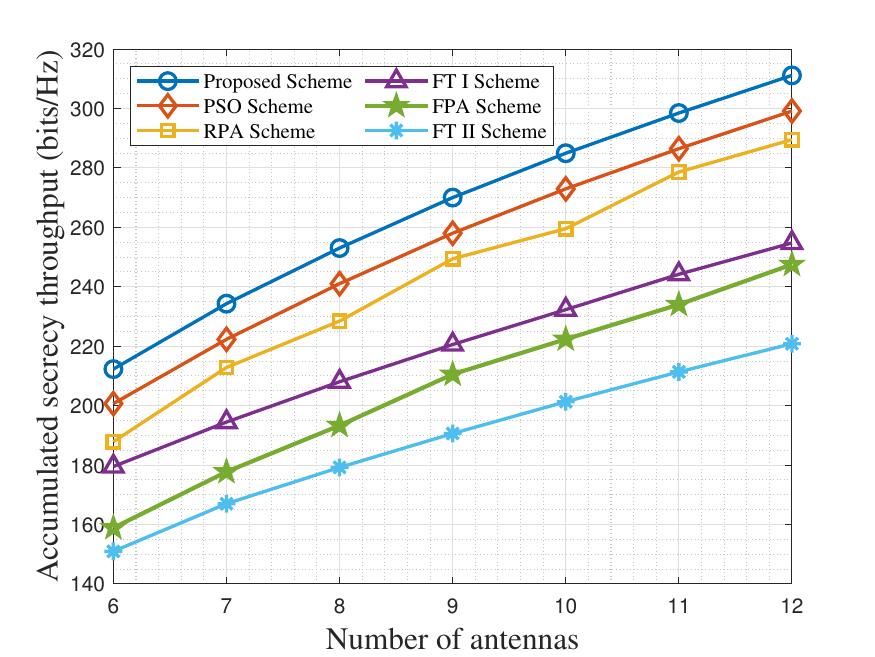}
\caption{Accumulated secrecy throughput versus the number of antennas.}
\label{number antenna}
\end{figure}

Fig. \ref{number antenna} illustrates the accumulated secrecy throughput of different schemes versus the number of UAV antennas. As the number of antennas increases, all schemes achieve higher accumulated secrecy throughput due to the increased spatial DoFs available for beamforming and interference suppression. The proposed scheme consistently outperforms the benchmark schemes over the considered antenna range. Specifically, its accumulated secrecy throughput increases from 212.31 bits/Hz with 6 antennas to 311.15 bits/Hz with 12 antennas, corresponding to an improvement of approximately 46.6\%. These results demonstrate the effectiveness of jointly optimizing the UAV trajectory, transmit beamforming, and FA positions in exploiting the additional spatial DoFs.

\begin{figure}[t]
\centering
\includegraphics[width=0.95\linewidth]{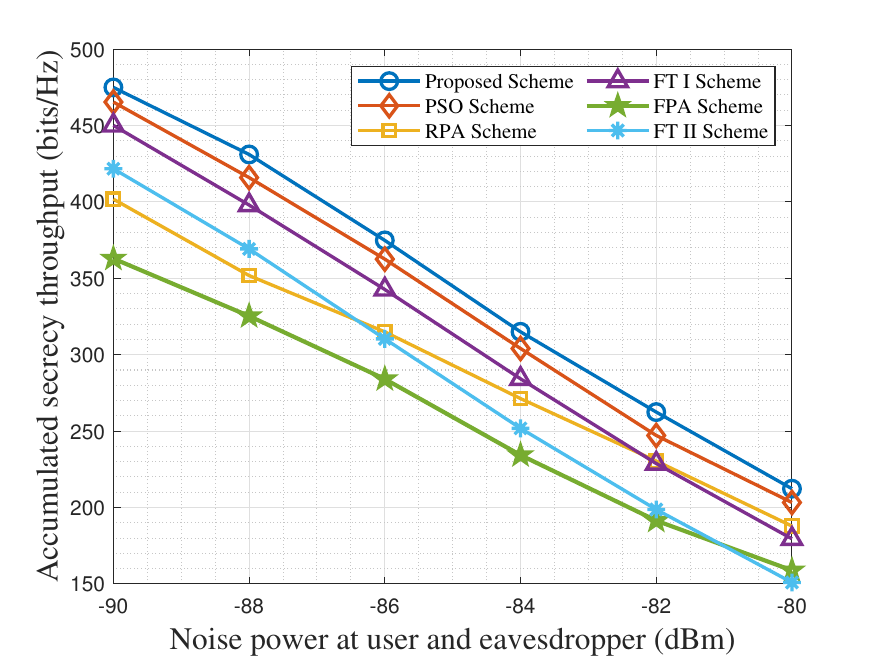}
\caption{Accumulated secrecy throughput versus the noise power at users and the eavesdropper.}
\label{noise}
\end{figure}

Fig. \ref{noise} illustrates the accumulated secrecy throughput of different schemes versus the noise power at both the legitimate users and the eavesdropper. As the noise power increases from $-90$ dBm to $-80$ dBm, the accumulated secrecy throughput of all schemes decreases. This trend suggests that, under the considered simulation settings, the degradation of legitimate communication links outweighs the benefit of reduced eavesdropping capability caused by the increased noise power. Nevertheless, the proposed scheme consistently achieves the highest accumulated secrecy throughput over the considered noise-power range. This result further demonstrates the effectiveness of jointly optimizing the UAV trajectory, transmit beamforming, and FA positions in enhancing secure communication performance.

\section{Conclusion}
In this paper, we have investigated an FA-enabled secure UAV communication system under uncertain eavesdropper locations. To enhance the expected accumulated secrecy throughput of multiple legitimate ground users, we jointly optimized the UAV trajectory, transmit beamforming, and FA positions. To address the non-convexity of the formulated problem, a sampling-based robust optimization framework was developed, followed by an AO-based algorithm to obtain a suboptimal solution to the resulting surrogate problem. Simulation results demonstrate that the proposed FA-enabled scheme achieves higher accumulated secrecy throughput than the conventional FPA scheme and other benchmark schemes. Future research may extend this work in several directions. First, it is valuable to investigate FA-enabled secure UAV networks in dynamic multi-UAV or swarm communication scenarios. Second, incorporating reconfigurable intelligent surfaces (RISs) into the proposed framework may further enhance spatial flexibility and secrecy performance. Third, the joint design of FA-enabled secure communication with mobile edge computing (MEC) and energy-efficient communication represents a promising avenue for integrating computation and security optimization. Furthermore, learning-based optimization methods, such as deep reinforcement learning (DRL), can be explored to achieve online trajectory and FA control under highly dynamic environments.

\begin{appendices}
\section{Detailed Derivations of $\mathbb{E} \big[ R_{m}[n] \big]$ and $\mathbb{E} \big[ R_{m, {\rm{E}}}[n] \big]$ }\label{proof1}
Since the instantaneous SINR depends on the random channel realizations, directly evaluating the expectation of the logarithmic rate expression is generally intractable. Following the statistical CSI-based approximation approaches in \cite{9500188,9034493,9355404}, we approximate the expected per-slot throughput by replacing the instantaneous desired signal and interference powers with their corresponding average values. Specifically, $\mathbb{E} \big[ R_{m}[n] \big]$ can be approximated as
\begin{equation}
\tilde{R}_{m}[n] = \left( \tau - t_{1}[n] \right) \log_{2} \left( 1 +  \tilde{\gamma}_{m}[n] \right), \label{proof1_1}
\end{equation}
where 
\begin{equation} 
\tilde{\gamma}_{m}[n] = \frac{  \mathbb{E} \{ | \mathbf{h}_{m}^{\mathsf{H}}[n] \mathbf{w}_{m}[n]  |^{2} \} }
{  \sum\limits_{ i = 1, i \neq m }^{ \mathcal{M} } 
\mathbb{E} \{ |  \mathbf{h}_{m}^{\mathsf{H}}[n] \mathbf{w}_{i}[n] |^{2} \} + \sigma^2_{m}}.
\end{equation} 

Based on the definition of the expectation operator $\mathbb{E}[\cdot]$, we can derive the following expression
\begin{align}
&\mathbb{E} \{ | \mathbf{h}_{m}^{\mathsf{H}}[n] \mathbf{w}_{i}[n]  |^{2} \} \nonumber \\
& = \mathbb{E} \left\{ \left| \left(\sqrt{ \zeta^{\rm{los}}_{m}[n] } \mathbf{\overline{h}}_{m}[n] + \sqrt{ \zeta^{\rm{nlos}}_{m}[n] } \mathbf{\tilde{h}}_{m}[n] \right)^{\mathsf{H}} \mathbf{w}_{i}[n]  \right|^{2} \right\} \nonumber \\
& = \zeta^{\rm{los}}_{m}[n] \left| \mathbf{\overline{h}}_{m}^{\mathsf{H}}[n]\mathbf{w}_{i}[n] \right|^{2} + \zeta^{\rm{nlos}}_{m}[n] \| \mathbf{w}_{i}[n] \|^{2}, \label{proof1_2}
\end{align}
where $\zeta^{\rm{los}}_{m}[n] = \frac{ \kappa_{m}[n] \beta_{m}[n] }{ \kappa_{m}[n] + 1 }$ and $\zeta^{\rm{nlos}}_{m}[n] = \frac{ \beta_{m}[n] }{ \kappa_{m}[n] + 1 }$.

By substituting \eqref{proof1_2} into \eqref{proof1_1}, we obtain the approximate ergodic rate expression in \eqref{lab2}. Similarly, the corresponding approximation for the eavesdropping rate in \eqref{lab3} can be obtained.

\section{ Proof of Negative Semi-Definiteness of $\mathbf{S}^{\rm{lb}}_{\tilde{m},i}[n]$ }\label{proof2}
We first define $\mathbf{1}_{K} = [1, 1, \dots, 1]^{\mathsf{T}} \in \mathbb{R}^{K \times 1}$, and introduce the diagonal matrix
\begin{equation}\small
\varkappa_{i}[n] = \mathsf{diag}\left( \sqrt{|w_{i,1}[n]| },\sqrt{|w_{i,2}[n]| }, \dots, \sqrt{|w_{i,K}[n]|}^{\mathsf{T}}\right) \in \mathbb{R}^{K \times K}.
\end{equation}
Thus, we can obtain $\mathsf{diag} \left(\mathbf{\tilde{w}}_{i}[n] \right) = \varkappa_{i}[n]\varkappa_{i}[n]$ and $\mathbf{\tilde{w}}_{i} [n] = \varkappa_{i}[n]\varkappa_{i}[n]\mathbf{1}_{K}$. 

Therefore, the quadratic form $\mathbf{x}^{\mathsf{T}}[n] \mathbf{S}^{\rm{lb}}_{\tilde{m},i}[n] \mathbf{x}[n]$ can be expressed as
\begin{align}
&\mathbf{x}^{\mathsf{T}}[n] \mathbf{S}^{\rm{lb}}_{\tilde{m},i}[n] \mathbf{x}[n] \nonumber \\
&= -2 \vartheta_{\tilde{m}}^{2}[n]  \mathbf{x}^{\mathsf{T}}[n]\left( \mu_{i}[n] \mathsf{diag} \left(\mathbf{\tilde{w}}_{i}[n] \right) -\mathbf{\tilde{w}}_{i} \mathbf{\tilde{w}}_{i}^{\mathsf{T}}[n] \right)\mathbf{x}[n], \nonumber \\
& = -2 \vartheta_{\tilde{m}}^{2}[n] \left( \mu_{i}[n] \| \varkappa_{i}[n]  \mathbf{x}[n] \|_2^2 -  | \mathbf{1}^{\mathsf{T}}_{K}  \varkappa_{i}[n]\varkappa_{i}[n] \mathbf{x}[n] |^{2}  \right) \nonumber \\
&= -2 \vartheta_{\tilde{m}}^{2}[n] \big( \| \varkappa_{i}[n] \mathbf{1}_{K} \|_2^2 \| \varkappa_{i}[n]  \mathbf{x}[n] \|_2^2 \nonumber \\
& \hspace{12em} - | \mathbf{1}^{\mathsf{T}}_{K}  \varkappa_{i}[n]\varkappa_{i}[n] \mathbf{x}[n] |^{2} \big). 
\end{align}
By the Cauchy-Schwarz inequality, which states that $\| \mathbf{u} \|_2^2 \| \mathbf{v} \|_2^2 \geq |\mathbf{u}^{\mathsf{T}} \mathbf{v}|^2$ for any two vectors $\mathbf{u}$ and $\mathbf{v}$ of the same dimension, we conclude that $\mathbf{x}^{\mathsf{T}}[n] \mathbf{S}^{\rm{lb}}_{\tilde{m},i}[n] \mathbf{x}[n] \leq 0$. Hence, the matrix $\mathbf{S}^{\rm{lb}}_{\tilde{m},i}[n]$ is negative semi-definite. This completes the proof.

\end{appendices}

\bibliographystyle{IEEEtran}
\bibliography{myref}

\end{document}